\documentclass[prd, onecolumn, nofootinbib]{revtex4-2}

\usepackage{fullpage}

\usepackage{amsfonts}
\usepackage{amsmath}
\usepackage{amssymb}

\usepackage{graphicx}

\usepackage{xcolor}

\usepackage[colorlinks=true,linkcolor=blue,citecolor=blue,urlcolor=blue]{hyperref}

\newcommand{\result}[1]{#1}

\begin{document}

\title{
Precision Requirements for Monte Carlo Sums within Hierarchical Bayesian Inference
}

\author{Reed Essick}
\email{ressick@perimeterinstitute.ca}
\affiliation{Perimeter Institute for Theoretical Physics, 31 Caroline Street North, Waterloo, Ontario, Canada, N2L 2Y5}

\author{Will M. Farr}
\email{will.farr@stonybrook.edu}
\affiliation{Center for Computational Astrophysics, Flatiron Institute, New York, NY 10010, USA}
\affiliation{Department of Physics and Astronomy, Stony Brook University, Stony Brook, NY 11794, USA}

\begin{abstract}
    Hierarchical Bayesian inference is often conducted with estimates of the target distribution derived from Monte Carlo sums over samples from separate analyses of parts of the hierarchy or from mock observations used to estimate sensitivity to a target population.
    We investigate requirements on the number of Monte Carlo samples needed to guarantee the estimator of the target distribution is precise enough that it does not affect the inference.
    We consider probabilistic models of how Monte Carlo samples are generated, showing that the finite number of samples introduces additional uncertainty as they act as an imperfect encoding of the components of the hierarchical likelihood. 
    Additionally, we investigate the behavior of estimators marginalized over approximate measures of the uncertainty, comparing their performance to the Monte Carlo point estimate.
    We find that correlations between the estimators at nearby points in parameter space are crucial to the precision of the estimate.
    Approximate marginalization that neglects these correlations will either introduce a bias within the inference or be more expensive (require more Monte Carlo samples) than an inference constructed with point estimates.
    We therefore recommend that hierarchical inferences with empirically estimated target distributions use point estimates.
\end{abstract}

\maketitle


\section{Introduction}

Hierarchical Bayesian inference plays a fundamental role in catalog science in that it allows one to combine information from many detected events to infer the properties of an entire population while self-consistently accounting for selection effects and measurement uncertainty within a survey.
What's more, the hierarchical structure of the inference allows the computation to proceed in a piece-meal fashion in which individual events are first analyzed separately and then ``stacked'' within the ``population inference.''
Thus, the input for hierarchical analyses is often sets of ``single-event parameter'' samples drawn from reference posterior distributions, each conditioned on data from a single event with a reference prior, and a set of detected simulated signals (injections) used to assess the catalog's sensitivity.
The inference for ``population-level parameters'' proceeds via importance sampling to obtain an estimate of a target distribution proportional to the desired population-level posterior.
Our motivation stems from the inference of the population of merging compact binaries with Gravitational-Wave (GW) observations.
See, e.g., Refs.~\cite{Mandel2016, Thrane2019, Vitale2020} for reviews.

However, as importance sampling is done with a finite set of Monte Carlo samples, it provides only an approximation to the actual target distribution.
Our goal is to investigate the behavior of the uncertainty in the population inference stemming from the finite Monte Carlo sample size.

Previous work investigated this in the context of Gaussian uncertainties on the selection function~\cite{Farr2019}, and the implications have been extensively used in the GW literature, e.g., \cite{Abbott2020, Abbott2021}.
In short, Ref.~\cite{Farr2019} defined a criterion for the minimum number of Monte Carlo samples used to empirically estimate the catalog's sensitivity by approximating the uncertainty from the Monte Carlo sum as a Gaussian.
Pragmatically, this suggests that parts of population parameters space that do not meet this criterion must be discarded as the target distribution is no longer valid.
Ref.~\cite{Talbot2020} used this to motivate other approximations to the selection function in order to avoid the need to discard any part of parameter space, but their framework still depends on an original set of samples and is limited by the uncertainty associated therewith.

Briefly, we find that precise inference is possible with a constant number of single-event posterior samples, regardless of the catalog's size, and a number of injections that scales linearly with the catalog's size (which is a constant number of injections \emph{per event}).
Target distributions that are marginalized over some approximations of Monte Carlo uncertainty require more samples to avoid biases.

Sec.~\ref{sec:formalism} defines our notation and lays out the basic approach to estimating the hierarchical likelihood with Monte Carlo sums.
We then consider general precision requirements for those Monte Carlo sums in Sec.~\ref{sec:general precision requirements}.
In particular, we examine the scaling of the number of Monte Carlo samples that are needed as the size of the catalog grows.
Sec.~\ref{sec:generative modeling} explores the uncertainty introduced by Monte Carlo Samples with a generative model of how data is produced within the catalog.
Sec.~\ref{sec:improved estimators} examines a few possible ways to marginalize our estimator for the hierarchical likelihood over (models of) the Monte Carlo uncertainty.
We conclude in Sec.~\ref{sec:discussion}.


\section{Formalism}
\label{sec:formalism}

We consider an hierarchical inference with a parametric population model described by the parameters $\Lambda$ and an overall rate $\mathcal{R}$.
Again, our motivation is the inference of population properties from catalogs of GW detections, but the conclusions hold more generally.

That is, we assume an inhomogeneous Poisson process with rate-density of signals given by
\begin{equation}
    \frac{d\mathcal{N}}{d\theta} = \mathcal{R}p(\theta|\Lambda)
\end{equation}
where $\theta$ refers to all single-event parameters and $\int d\theta\, p(\theta|\Lambda) = 1 \ \forall \ \Lambda$.
As such, the joint probability of observing data $\{D_i\}$ for $N$ events, $\Lambda$, and $\mathcal{R}$ with a prior $p(\Lambda, \mathcal{R})$ is
\begin{equation} \label{eq:target}
    p(\{D_i\}, \Lambda, \mathcal{R}) = p(\Lambda, \mathcal{R}) \mathcal{R}^N e^{-\mathcal{R} \mathcal{E}} \prod\limits_i^N \mathcal{Z}_i
\end{equation}
where
\begin{align}
    \mathcal{Z}_i(\Lambda)
        & \equiv p(D_i|\Lambda) = \int d\theta\, p(\theta|\Lambda) p(D_i|\theta) & \text{(single-event evidence)} \label{eq:single-event evidence} \\
    \mathcal{E}(\Lambda)
        & \equiv P(\mathrm{det}|\Lambda) = \int d\theta\, p(\theta|\Lambda) P(\mathrm{det}|\theta) & \text{(selection function)} \label{eq:selection function} \\
    P(\mathrm{det}|\theta)
        & \equiv \int\limits_{\{D\}_\mathrm{det}} dD\, p(D|\theta) & \text{(probability of detection)} \label{eq:detection probability}
\end{align}
and $\{D\}_\mathrm{det}$ refers to the set of detectable data.
This is usually defined in terms of a selection threshold, such as a minimum signal-to-noise ratio or a maximum false alarm probability.
It is also possible to jointly infer the parameters of individual events at the same time as the population parameters.
However, this can always be decomposed into an inference for the population parameters alone and then an inference for the single-event parameters conditioned on the population parameters.
See, e.g., Ref.~\cite{Essick2021} for more details.
As such, we focus on the population inference.

These integrals are generally intractable analytically, and we instead approximate them with Monte Carlo sums following the rule
\begin{equation}
    \int dx\, p(x) f(x) \approx \frac{1}{m}\sum\limits_j^m f(x_j)
\end{equation}
with $m$ independently and identically distributed (i.i.d.) Monte Carlo samples drawn from
\begin{equation}
    x_j \sim p(x)
\end{equation}
Specifically, we define an estimator for each single-event evidence using $m_i$ samples drawn from a reference posterior (with prior parameterized by $\Lambda_0$) separately for each event $i$
\begin{equation} \label{eq:Zhat}
    \frac{\hat{\mathcal{Z}}_i}{p(D_i|\Lambda_0)} \equiv \frac{1}{m_i} \sum\limits_j^{m_i} \frac{p(\theta_j|\Lambda)}{p(\theta_j|\Lambda_0)}
\end{equation}
with
\begin{equation}
    \theta_j \sim p(\theta|D_i,\Lambda_0) = \frac{p(D_i|\theta)p(\theta|\Lambda_0)}{p(D_i|\Lambda_0)}
\end{equation}
Similarly, we approximate the selection function with sets of $m$ detected simulated signals (injections) out of a set of $M$ drawn from a reference distribution
\begin{equation} \label{eq:Ehat}
    \hat{\mathcal{E}} \equiv \frac{1}{M}\sum\limits_k^m \frac{p(\theta_k|\Lambda)}{p(\theta_k|\mathrm{draw})}
\end{equation}
with
\begin{equation}
    \theta_k \sim p(\theta|\mathrm{det},\mathrm{draw}) = \frac{p(\theta|\mathrm{draw})P(\mathrm{det}|\theta)}{P(\mathrm{det}|\mathrm{draw})}
\end{equation}
and we approximate
\begin{equation}
    \hat{P}(\mathrm{det}|\mathrm{draw}) \approx \frac{m}{M}
\end{equation}
It is our goal to understand how uncertainty in these estimators (Eqs.~\ref{eq:Zhat} and~\ref{eq:Ehat}) impacts our estimator for the target distribution (Eq.~\ref{eq:target}).


\subsection{Rate-Marginalized Target Distribution}
\label{sec:rate-marginalized target distribution}

It is sometimes convenient to analytically marginalize out the rate ($\mathcal{R}$) with an appropriate choice of prior.
A good choice is the scaled $\Gamma$-distribution (also called the Erlang distribution)
\begin{equation}
    p(\mathcal{R}) = \frac{\mathcal{R}^{n-1} e^{-\mathcal{R}/r}}{r^n \Gamma(n)}
\end{equation}
described by two parameters: $n$ and $r$.
With this in hand, one can marginalize over $\mathcal{R}$ analytically, obtaining
\begin{equation}
    p(\{D_i\}|\Lambda) \propto \left(\mathcal{E} + \frac{1}{r}\right)^{-(N+n)} \left[\prod\limits_i^N Z_i\right]
\end{equation}
where the constant of proportionality does not depend on $\Lambda$ and therefore will not affect the shape of the posterior for $\Lambda$.
A common choice is $r \rightarrow \infty$ and $n = 0$ so that $p(\mathcal{R}) \sim 1/\mathcal{R}$.
Note that $r$ acts to stabilize the target distribution against very small $\mathcal{E}$, which would otherwise cause the rate-marginalized likelihood to diverge.
In effect, then, $r$ sets an upper limit for the target distribution.\footnote{
The limit $\mathcal{E} \rightarrow 0$ corresponds to $\mathcal{R} \rightarrow \infty$ if the expected number of detections ($\mathcal{E}\mathcal{R}$) is constant.
It is because $r$ limits the prior for very large $\mathcal{R}$ that it then limits the marginal likelihood for very small $\mathcal{E}$.
}

We will focus on the behavior of an estimator for this rate-marginalized likelihood constructed from Monte Carlo sums.
That is, we consider
\begin{equation}
    \hat{p}(\{D_i\}|\Lambda) \propto \left( \frac{1}{r} + \frac{1}{M}\sum\limits_k^m \frac{p(\theta_k|\Lambda)}{p(\theta_k|\mathrm{draw})} \right)^{-(N+n)} \left[\prod\limits_i^N \left( \frac{1}{m_i}\sum\limits_j^{m_i} \frac{p(\theta_j|\Lambda)}{p(\theta_j|\Lambda_0)} \right) \right]
\end{equation}
Note that this immediately suggests $r \sim M$, as one does not expect to be able to measure the selection function more precisely than this with $M$ injections.
Alternatively, one may consider physical priors for $r$, such as an upper limit given by some fraction of the peak star formation rate throughout the history of the universe~\cite{Madau:2014bja, Madau:2016jbv}.


\subsection{Time-Varying Survey Sensitivity, Waveform Uncertainty, and Calibration Error}
\label{sec:waveform uncertainty}

Although we will not explore them in any great detail, it is worth mentioning a few things that can affect a survey.
Specifically,
\begin{itemize}
    \item the sensitivity of detectors may change over time,
    \item there may be errors in the models used to approximate signals (waveform errors), and
    \item there may be errors in the model of the detector response which change the way signals appear in the detector (calibration errors).
\end{itemize}
We proceed under the assumption that the detector sensitivity in the vicinity of each detection is stationary.
See, e.g., Refs.~\cite{Chatziioannou:2019zvs, Littenberg:2014oda, Talbot:2020auc, Biscoveanu:2020kat, Zackay:2019kkv} for more discussion.
This means that we need not be concerned the impact of time-varying detector sensitivity within our models of the single-event evidences in Eq.~\ref{eq:target}.
The impact of waveform and calibration uncertainty on single-event evidences, and the best way to deal with them, is still an open question.
We will not consider them in more detail here.
See, e.g., Refs.~\cite{Essick2022, Purrer2019, Ashton:2021yum} for more discussion.

We are then concerned primarily with the selection function $\mathcal{E}$.
Eq.~\ref{eq:selection function} shows that the sensitivity to a population is an integral over the population, with the probability of detection given by an integral over detectable data (Eq.~\ref{eq:detection probability}).
As written, the latter is a bit of an oversimplification.
In particular, the catalog's sensitivity is an average over the entire duration of the experiment
\begin{equation}
    P(\mathrm{det}|\theta) = \int dt\, p(t) P(\mathrm{det}|\theta,t)
\end{equation}
where $p(t)$ describes the prior for when signals occur (almost always taken to be uniform throughout the duration of the experiment).
With a model of how the survey's sensitivity varies through time, one can then account for time-dependent sensitivity directly.
This is typically done by injecting simulated signals throughout the duration of the experiment so that the marginalization over time is implicitly included within the Monte Carlo sum.

Additionally, if one acknowledges that models for astrophysical signals and the detector response are probabilistic (they are not known with perfect certainty), then one should likewise marginalize over the associated uncertainty
\begin{equation}
    P(\mathrm{det}|\theta, t) = \int \mathcal{D}h \, p(h|\theta) P(\mathrm{det}|h, t)
\end{equation}
where $\mathcal{D}h\, p(h|\theta)$ represent the measure with respect to the functional degrees of freedom in the way a signal with parameters $\theta$ appears in the detector.
In practice, this could be done by drawing injected waveforms from $p(h|\theta)$.
Again, the marginalization would automatically be included within the Monte Carlo summation.
Alternatively, Ref.~\cite{Essick2022} studies situations in which the marginalization may be carried out analytically, although further study is needed before that may be practical.


\section{General Precision Requirements}
\label{sec:general precision requirements}

When constructing general precision requirements for our estimate of the target distribution, our goal is to reconstruct the true target distribution with only small uncertainty.
We define this as the ability to faithfully estimate the moments of population parameters taken with respect to the true target distribution from the estimator.
That is, statistics like
\begin{equation}
    \mathrm{E}[\Lambda]_{\hat{p}(\Lambda|\{D_i\})} \equiv \int d\Lambda \hat{p}(\Lambda|\{D_i\}) \Lambda
\end{equation}
must always be close to the true value obtained with an integral over $p(\Lambda|\{D_i\})$ instead of $\hat{p}(\Lambda|\{D_i\})$.

An immediate possibility is to require the variance of $\hat{p}$ under Monte Carlo uncertainty to be small compared to the expected value (squared).
That is,\footnote{Here and throughout, we denote moments taken with respect to different Monte Carlo realizations with the subscript ``MC.''}
\begin{equation}
    \mathrm{Var}[\hat{p}]_\mathrm{MC} \ll \mathrm{E}[\hat{p}]_\mathrm{MC}^2 \label{eq:criterion 1}
\end{equation}
By design, this accounts for the arbitrary scale of $\hat{p}$.
This is desirable because we are only interested in the shape of the target distribution.
An alternate way to express Eq.~\ref{eq:criterion 1} is to require the variance of $\ln \hat{p}$ to be small.
This criterion will also be independent of the overall scale of $\hat{p}$.
In fact, one may approximate the variance of $\ln \hat{p}$ as
\begin{equation}
    \mathrm{Var}[\ln \hat{p}]_\mathrm{MC} \approx \frac{\mathrm{Var}[\hat{p}]_\mathrm{MC}}{\mathrm{E}[\hat{p}]_\mathrm{MC}^2}
\end{equation}
We see that this is the same statement as Eq.~\ref{eq:criterion 1}.

However, we do not actually care about the value of $\ln p$ at any individual point in parameter space.
Instead, we are only interested in the difference in $\ln p$ between points in parameter space; the difference $\Delta \ln p = \ln p(\Lambda_a|\{D_i\}) - \ln p(\Lambda_b|\{D_i\})$ determines the shape of the distribution.
That is, correlations between the estimator at different points in population parameter space may improve our estimates of the target distribution's shape.
One may then consider the requirement that $\mathrm{Var}[\Delta \ln \hat{p}]_\mathrm{MC}$ between any two fixed points in parameter space be small.

While it can be difficult to estimate the variance of the log of Monte Carlo sums, it is relatively straightforward to estimate the variance of the sums themselves.
We consider the variance of the log of the estimator for the rate-marginalized target distribution
\begin{equation}
    \mathrm{Var}[\Delta \ln \hat{p}]_\mathrm{MC} = (N + n)^2 \mathrm{Var}\left[\Delta \ln \left(\hat{\mathcal{E}} + r^{-1}\right)\right]_\mathrm{MC} + \sum_i^N \mathrm{Var}[\Delta \ln \hat{\mathcal{Z}}_i]_\mathrm{MC}
\end{equation}
and approximate each term as follows
\begin{align}
    \mathrm{Var}[\Delta \ln \hat{\mathcal{Z}}]_\mathrm{MC}
        & \approx \frac{\mathrm{Var}[\hat{\mathcal{Z}}(\Lambda_a)]_\mathrm{MC}}{\mathrm{E}[\hat{\mathcal{Z}}(\Lambda_a)]_\mathrm{MC}^2}
            + \frac{\mathrm{Var}[\hat{\mathcal{Z}}(\Lambda_b)]_\mathrm{MC}}{\mathrm{E}[\hat{\mathcal{Z}}(\Lambda_b)]_\mathrm{MC}^2}
            - \frac{2\mathrm{Cov}[\hat{\mathcal{Z}}(\Lambda_a), \hat{\mathcal{Z}}(\Lambda_b)]_\mathrm{MC}}{\mathrm{E}[\hat{\mathcal{Z}}(\Lambda_a)]_\mathrm{MC}\mathrm{E}[\hat{\mathcal{Z}}(\Lambda_b)]_\mathrm{MC}} \label{eq:var delta lnz} \\
    \mathrm{Var}\left[\Delta \ln \left( \hat{\mathcal{E}} + r^{-1} \right)\right]_\mathrm{MC}
        & \approx \frac{\mathrm{Var}[\hat{\mathcal{E}}(\Lambda_a)]_\mathrm{MC}}{(\mathrm{E}[\hat{\mathcal{E}}(\Lambda_a)]_\mathrm{MC} + r^{-1})^2}
            + \frac{\mathrm{Var}[\hat{\mathcal{E}}(\Lambda_b)]_\mathrm{MC}}{(\mathrm{E}[\hat{\mathcal{E}}(\Lambda_b)]_\mathrm{MC} + r^{-1})^2} \nonumber \\
        & \quad \quad
            - \frac{2\mathrm{Cov}[\hat{\mathcal{E}}(\Lambda_a), \hat{\mathcal{E}}(\Lambda_b)]_\mathrm{MC}}{(\mathrm{E}[\hat{\mathcal{E}}(\Lambda_a)]_\mathrm{MC} + r^{-1})(\mathrm{E}[\hat{\mathcal{Z}}(\Lambda_b)]_\mathrm{MC} + r^{-1})} \label{eq:var delta e}
\end{align}
Appendix~\ref{sec:estimators} provides expressions for each of these terms as well as ways to estimate them from a single realization of Monte Carlo samples.
The most important result is that both Eq.~\ref{eq:var delta lnz} and~\ref{eq:var delta e} scale inversely with the size of the Monte Carlo sample set.
That is, for any two fixed points in parameter space, $M$ injections, and $\{m_i\}$ single-event posterior samples
\begin{equation}
    \mathrm{Var}[\Delta \ln \hat{p}]_\mathrm{MC} \approx \mathcal{O}\left(\frac{(N+n)^2}{M}\right) + \sum\limits_i^N \mathcal{O}\left(\frac{1}{m_i}\right) \label{eq:fixed param var scaling}
\end{equation}
As the catalog grows, then, this suggests we need
\begin{gather}
    M \sim \mathcal{O}(N^2) \label{eq:sufficient M scaling} \\
    m_i \sim \mathcal{O}(N) \label{eq:sufficient m_i scaling}
\end{gather}
in order to maintain $\mathrm{Var}[\Delta \ln \hat{p}]_\mathrm{MC} \ll 1$.
These scalings are not an issue for relatively small catalogs, but they could make precise inference over larger catalogs impractical.

However, while this should be sufficient to accurately resolve the moments of the population parameters \textit{a posteriori}, it may not be necessary.
That is, Eq.~\ref{eq:fixed param var scaling} sets a requirement to accurately resolve the shape of the posterior between any two fixed points in parameter space.
Instead, we only care about the shape of the distribution where there is nontrivial posterior support, and that region shrinks as the catalog grows.
That is, we care about bounding
\begin{equation}
    \int d\Lambda p(\Lambda|\{D_i\}) \mathrm{Var}[\ln \hat{p}(\Lambda|\{D_i\}) - \ln \hat{p}(\bar{\Lambda}|\{D_i\})]_\mathrm{MC} \ll 1 \label{eq:best criterion}
\end{equation}
where we have selected as a reference the mean population parameters (with respect to the true posterior)
\begin{equation}
    \bar{\Lambda} \equiv \int d\Lambda p(\Lambda|\{D_i\}) \Lambda
\end{equation}
Furthermore, if we assume $p(\Lambda|\{D_i\})$ has support only near $\bar{\Lambda}$, then we can approximate
\begin{equation}
    \mathrm{Var}[\ln \hat{p}(\Lambda|\{D_i\}) - \ln \hat{p}(\bar{\Lambda}|\{D_i\})]_\mathrm{MC} \sim \left( \mathcal{O}\left(\frac{(N+n)^2}{M}\right) + \sum\limits_i^N \mathcal{O}\left(\frac{1}{m_i}\right) \right) \mathcal{O}(\Lambda-\bar{\Lambda})^2 \label{eq:quadratic scaling}
\end{equation}
where the terms in the square brackets only depend on integrals of the target distribution and its derivatives with respect to $\Lambda$ at $\bar{\Lambda}$.
See Appendix~\ref{sec:estimators}.
Again, a key result it that the variance of estimators scale inversely with the number of Monte Carlo samples.
But, additionally, the variance scales quadratically with the separation in parameter space.

Inserting Eq.~\ref{eq:quadratic scaling} into Eq.~\ref{eq:best criterion}, we obtain
\begin{align}
    \int d\Lambda p(\Lambda|\{D_i\}) \mathrm{Var}[\ln \hat{p}(\Lambda|\{D_i\}) - \ln \hat{p}(\bar{\Lambda}|\{D_i\})]_\mathrm{MC}
        & \approx \frac{1}{2} \, g_{ij} \, \int d\Lambda p(\Lambda|\{D_i\}) (\Lambda - \bar{\Lambda})_i (\Lambda - \bar{\Lambda})_j \nonumber \\
        & = \frac{1}{2} \, g_{ij} \, \mathrm{Cov}[\Lambda_i, \Lambda_j]_{p(\Lambda|\{D_i\})}
\end{align}
where the covariance is taken with respect to the true posterior and $g_{ij}$ represents a metric for how much the variance grows as a function of the parameter separation.
If the posterior is relatively peaked, then we may approximate the covariance with the inverse of the Fisher Information matrix.
In this case, we expect $\mathrm{Cov}[\Lambda_i, \Lambda_j]_{p(\Lambda|\{D_i\})} \sim 1/N$.
Therefore, Eq.~\ref{eq:best criterion} scales as
\begin{equation}
    \int d\Lambda p(\Lambda|\{D_i\}) \mathrm{Var}[\Delta \ln \hat{p}]_\mathrm{MC}
        \approx \left(\mathcal{O}\left(\frac{(N+n)^2}{M}\right) + \sum_i^N \mathcal{O}\left(\frac{1}{m_i}\right)\right) \mathcal{O}\left(\frac{1}{N}\right) \label{eq:best scaling}
\end{equation}
which suggests we only need
\begin{gather}
    M \sim \mathcal{O}(N) \label{eq:best criterion M scaling} \\
    m_i \sim \mathrm{constant} \label{eq:best criterion m_i scaling}
\end{gather}
in order to perform precise inference.\footnote{
It might be tempting to assert a precision requirement like $\mathrm{Var}[\Delta \ln \hat{p}]_\mathrm{MC} \ll \mathrm{E}[\Delta \ln \hat{p}]_\mathrm{MC}^2$.
However, this is ``too loose.''
For example, we expect $\mathrm{E}[\Delta \ln \hat{p}]_\mathrm{MC} \sim \mathcal{O}(N)$ and the resulting requirement suggests that we need fewer single-event Monte Carlo samples per detection as the catalog grows.
See Fig.~\ref{fig:including-selection-effects} for a demonstration of the breakdown of estimates of the selection function.
}
We explore the behavior of these criteria quantitatively with an analytically tractable toy model in Sec.~\ref{sec:univariate-gaussian toy model}.

We note that Ref.~\cite{Farr2019} came to a similar conclusion by approximating the marginalization of the selection function over Monte Carlo uncertainty with a Gaussian and then examining derivatives of the resulting log-target function (the size of a correction to the likelihood from the marginalization scales as $\mathcal{O}(N/N_\mathrm{eff})$).
While the analysis in Ref.~\cite{Farr2019} only holds when the approximate marginalization is valid, and there are other concerns with target distributions marginalized over approximate measures of Monte Carlo uncertainty, our conclusion should hold more generally.
We will discuss this in more detail in Sec.~\ref{sec:improved estimators}.


\subsection{Univariate-Gaussian Toy Model}
\label{sec:univariate-gaussian toy model}

To study these considerations quantitatively, we investigate an analytically tractable toy model.
Specifically, we assume each detected event is characterized by a single number ($x$).
We assume a Gaussian population
\begin{equation}
    p(x|\mu_\mathrm{pop}, \sigma_\mathrm{pop}) = \frac{1}{\sqrt{2\pi\sigma_\mathrm{pop}^2}} \exp \left( -\frac{(x - \mu_\mathrm{pop})^2}{2\sigma_\mathrm{pop}^2} \right) 
\end{equation}
We also assume that detector noise is Gaussian so that the observed parameter for each event ($x_\mathrm{obs}$) is Gaussian distributed and the single-event evidence ($\mathcal{Z}$) is Gaussian
\begin{align}
    p(x_\mathrm{obs}|x, \sigma_{obs})
        & = \left( 2\pi\sigma_\mathrm{obs}^2 \right)^{-1/2} \exp \left( -\frac{(x_\mathrm{obs} - x)^2}{2\sigma_\mathrm{obs}} \right) \\
    \mathcal{Z} \equiv p(x_\mathrm{obs}|\mu_\mathrm{pop}, \sigma_\mathrm{pop}, \sigma_\mathrm{obs})
        & = \left( 2\pi(\sigma_\mathrm{obs}^2 + \sigma_\mathrm{pop}^2) \right)^{-1/2} \exp \left( -\frac{(x_\mathrm{obs} - \mu_\mathrm{pop})^2}{2(\sigma_\mathrm{obs}^2 + \sigma_\mathrm{pop}^2)} \right)
\end{align}
Finally, we assume a Gaussian detection probability conditioned on $x_\mathrm{obs}$.
Integrals over the single-event likelihood and population yield a Gaussian for the selection function ($\mathcal{E}$)
\begin{align}
    P(\mathrm{det}|x_\mathrm{obs})
        & = \exp\left( -\frac{(x_\mathrm{obs} - \mu_\mathrm{det})^2}{2\sigma_\mathrm{det}^2} \right) \\
    P(\mathrm{det}|x, \sigma_\mathrm{obs})
        & = \int dx_\mathrm{obs}\, p(x_\mathrm{obs}|x, \sigma_\mathrm{obs}) P(\mathrm{det}|x_\mathrm{obs}) \nonumber \\
        & = \left( \frac{\sigma_\mathrm{det}^2}{\sigma_\mathrm{det}^2 + \sigma_\mathrm{obs}^2} \right)^{1/2} \exp \left( - \frac{(x - \mu_\mathrm{det})^2}{2(\sigma_\mathrm{det}^2 + \sigma_\mathrm{obs}^2)} \right) \\
    \mathcal{E} \equiv P(\mathrm{det}|\mu_\mathrm{pop}, \sigma_\mathrm{pop}, \sigma_\mathrm{obs}) 
        & = \int dx\, p(x|\mu_\mathrm{pop}, \sigma_\mathrm{pop}) P(\mathrm{det}|x, \sigma_\mathrm{obs}) \nonumber \\
        & = \left( \frac{\sigma_\mathrm{det}^2}{\sigma_\mathrm{det}^2 + \sigma_\mathrm{obs}^2 + \sigma_\mathrm{pop}^2} \right)^{1/2} \exp \left( - \frac{(\mu_\mathrm{pop} - \mu_\mathrm{det})^2}{2(\sigma_\mathrm{det}^2 + \sigma_\mathrm{obs}^2 + \sigma_\mathrm{pop}^2)} \right)
\end{align}
Because these expressions are all analytic, we can easily compute the exact target distribution.
We then generate mock catalogs of $N$ detections based in this generative model.
With the same set of observed events, we compare the behavior of an estimator for the rate-marginalized target distribution constructed from Monte Carlo sums to the analytic distribution.

Specifically, we are interested in estimators of the moments of the population posterior.
We focus on
\begin{equation}
    \mathrm{E}[\mu_\mathrm{pop}]_{\hat{p}} \equiv \int d\mu_\mathrm{pop} d\sigma_\mathrm{pop}\, \hat{p}(\mu_\mathrm{pop}, \sigma_\mathrm{pop}|\{x_i\}) \mu_\mathrm{pop}
\end{equation}
for simplicity and compute the distributions of this moment under Monte Carlo uncertainty by directly simulating many Monte Carlo realizations.
In what follows, we simulate catalogs with true population parameters \result{$(\mu_\mathrm{pop}, \sigma_\mathrm{pop}) = (0, 2)$}, single-event noise parameter \result{$\sigma_\mathrm{obs} = 1.0$}, and, when applicable, selection function parameters \result{$(\mu_\mathrm{det}, \sigma_\mathrm{det}) = (1, 3)$}.
We focus on the rate-marginalized likelihood with \result{$p(\mathcal{R}) \sim 1/\mathcal{R}$} and impose uniform priors with support over \result{$(-5 \leq \mu_\mathrm{pop} \leq +5)$} and \result{$(1 \leq \sigma_\mathrm{pop} \leq 10)$}.

\begin{figure}
    \begin{minipage}{0.49\textwidth}
        \begin{center}
            \large{Exact \quad \quad \quad} \\
            \vspace{0.1cm}
            \includegraphics[width=1.0\textwidth, clip=True, trim=1.7cm 0.2cm 1.3cm 1.10cm]{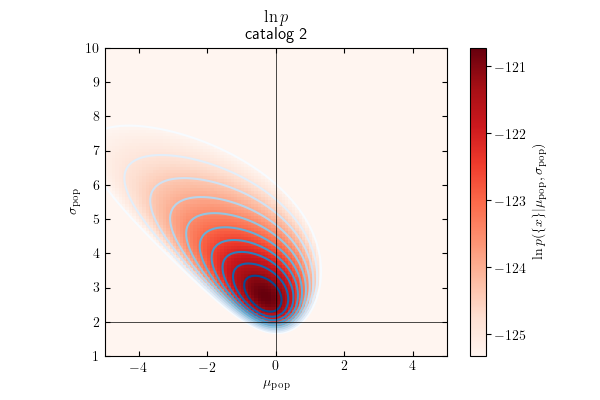}
        \end{center}
    \end{minipage}
    \begin{minipage}{0.49\textwidth}
        \begin{center}
            \large{Estimate \quad \quad \quad} \\
            \vspace{0.1cm}
            \includegraphics[width=1.0\textwidth, clip=True, trim=1.7cm 0.2cm 1.3cm 1.10cm]{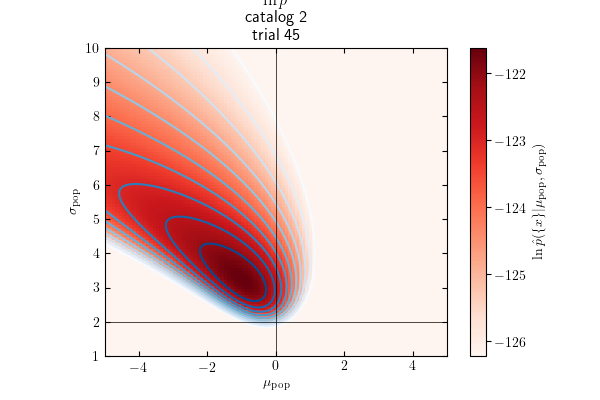}
        \end{center}
    \end{minipage}
    \caption{
        Exact and estimated target distribution for a catalog with \result{64} events.
        The estimated target distribution is constructed with \result{100} single-event posterior samples per event and \result{1000} injections.
        Fig.~\ref{fig:including-selection-effects} shows that the estimate of the target distribution begins to introduce errors comparable to the width of the exact posterior with this number of injections for this catalog size.
        Notice that both the maximum and the width of the estimate for the target distribution differ from the exact result, but both distributions have the same general features.
        In particular, the estimate of the target distribution is still a smooth function of the population parameters.
    }
    \label{fig:example target distribs}
\end{figure}

Fig.~\ref{fig:example target distribs} shows an example catalog with \result{64} events.
It shows the exact target distribution as well as an example of the estimator for the target distribution obtained using Monte Carlo sums.
In what follows, we first examine the requirement for the number of single-event posterior samples (Sec.~\ref{sec:single-event posterior samples}) in the context of catalogs where all events are detectable.
Sec.~\ref{sec:injections} then considers the requirement for the number of injections.


\subsubsection{Impact of the Number of Single-Event Posterior Samples}
\label{sec:single-event posterior samples}

We first examine the scaling required for the number of single-event posterior samples per event.
To do this, we simulate catalogs of different sizes.
For each catalog, we compute the exact target distribution as well as the estimate of the target distribution for each of $\mathcal{O}(100)$ different realizations of the single-event posterior sample sets.
This allows us to directly measure the variability in the estimate of the population posterior relative to the exact result.

\begin{figure*}
    \begin{minipage}{0.49\textwidth}
        \begin{center}
            \large{$\quad \quad \quad m_i = 100$} \\
            \includegraphics[width=1.0\textwidth]{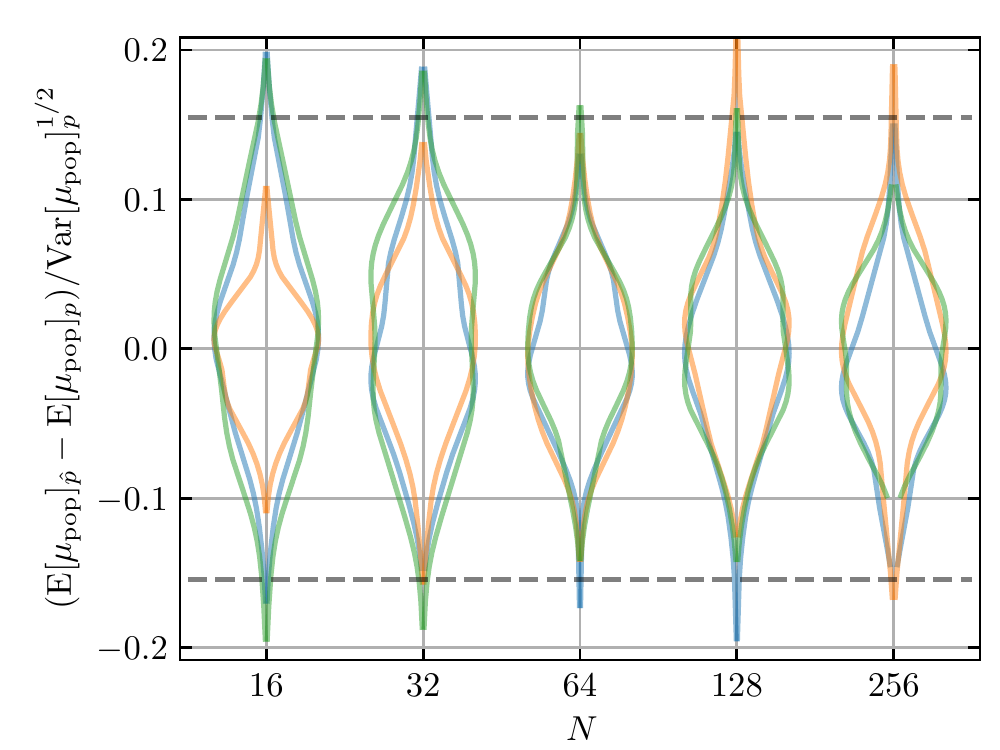}
        \end{center}
    \end{minipage}
    \begin{minipage}{0.49\textwidth}
        \begin{center}
            \large{$\quad \quad \quad m_i = 100 N$} \\
            \includegraphics[width=1.0\textwidth]{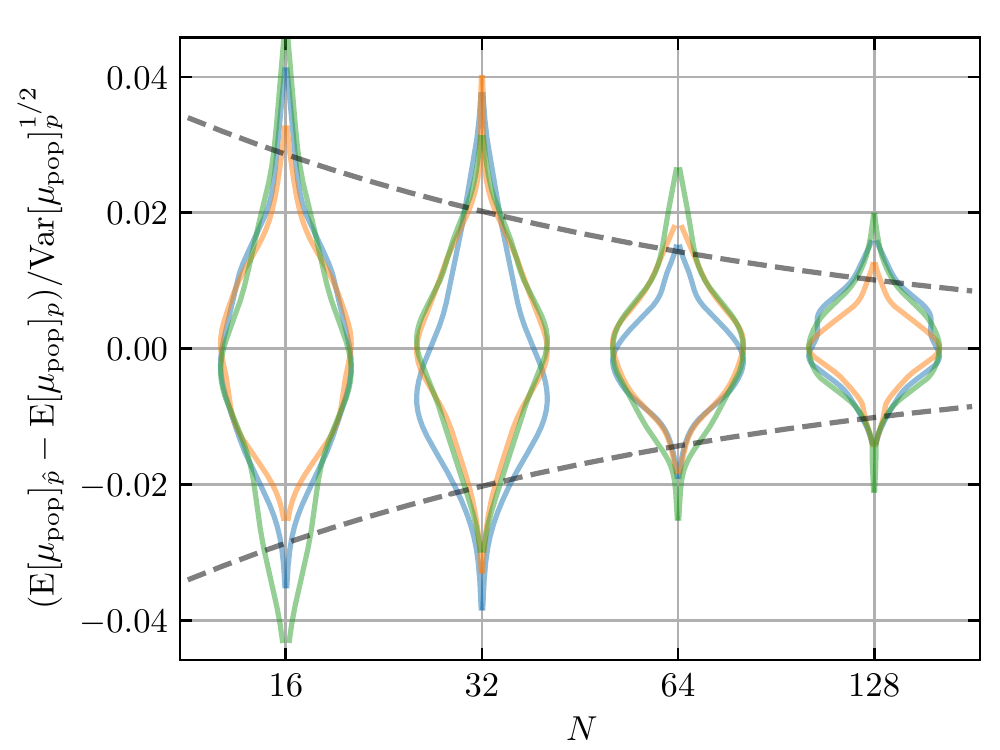}
        \end{center}
    \end{minipage}
    \caption{
        Variation in the mean \textit{a posteriori} derived from the empirical target distribution ($\mathrm{E}[\mu_\mathrm{pop}]_{\hat{p}}$) with respect to that derived from the exact distribution ($\mathrm{E}[\mu_\mathrm{pop}]_{p}$) scaled by the uncertainty from the exact distribution ($\mathrm{Var}[\mu_\mathrm{pop}]_{p}^{1/2}$), all using the same set of detected events.
        If Monte Carlo uncertainty is not to impact the inference, we require $\mathrm{E}[\mu_\mathrm{pop}]_{\hat{p}}$ to be tightly distributed around $\mathrm{E}[\mu_\mathrm{pop}]_{p}$ in comparison to $\mathrm{Var}[\mu_\mathrm{pop}]_{p}^{1/2}$.
        These simulations consider catalogs in which every event is detectable ($\mathcal{E} = 1$), which allows us to only consider the impact of single-event posterior samples.
        We consider cases in which (\emph{left}) the number of Monte Carlo samples per event is constant regardless of the catalog size and (\emph{right}) the number of Monte Carlo samples per event scales linearly with the catalog size.
        Each color represent a different realization of the catalog (different detected events), and the distributions show the variability from different realizations of the Monte Carlo samples using the same set of detected events.
        Dashed lines represent the expected scaling based on the considerations in Sec.~\ref{sec:general precision requirements}.
        In agreement with Eq.~\ref{eq:best criterion m_i scaling}, we only need a constant number of single-event samples per event regardless of the catalog size to maintain precise inference.
    }
    \label{fig:everything-is-detectable}
\end{figure*}

Fig.~\ref{fig:everything-is-detectable} summarizes our main conclusion.
That is, it shows the difference between the mean $\mu_\mathrm{pop}$ \textit{a posteriori} computed with the Monte Carlo estimate of the target distribution ($\mathrm{E}[\mu_\mathrm{pop}]_{\hat{p}}$) and the same mean computed with the exact target distribution ($\mathrm{E}[\mu_\mathrm{pop}]_p$).
We scale this difference by the uncertainty in $\mu_\mathrm{pop}$ from the exact target distribution ($\mathrm{Var}[\mu_\mathrm{pop}]_p^{1/2}$).
In order for the population inference to be precise, this difference must be tightly distributed around zero in comparison to the uncertainty in the posterior.
Fig.~\ref{fig:everything-is-detectable} shows two possible scalings: the number of Monte Carlo samples per event is either constant or proportional to the size of the catalog.
These correspond to the scalings suggested in Eq.~\ref{eq:best criterion m_i scaling} and Eq.~\ref{eq:sufficient m_i scaling}, respectively.

As is readily apparent from Fig.~\ref{fig:everything-is-detectable}, scaling the number of posterior samples per event with the size of the catalog is sufficient for precise inference, but it is not necessary.
That is, the relative uncertainty from different Monte Carlo realizations shrinks as the size of the catalog grows.
Instead, if we keep the number of samples per event the same as the catalog grows, we also see that the relative uncertainty from different Monte Carlo realizations is constant.


\subsubsection{Impact of the Number of Injections}
\label{sec:injections}

\begin{figure*}
    \begin{minipage}{0.49\textwidth}
        \begin{center}
            \large{$\quad \quad \quad m_i = 100$} \\
            \large{$\quad \quad \quad M = 1000$} \\
            \includegraphics[width=1.0\textwidth]{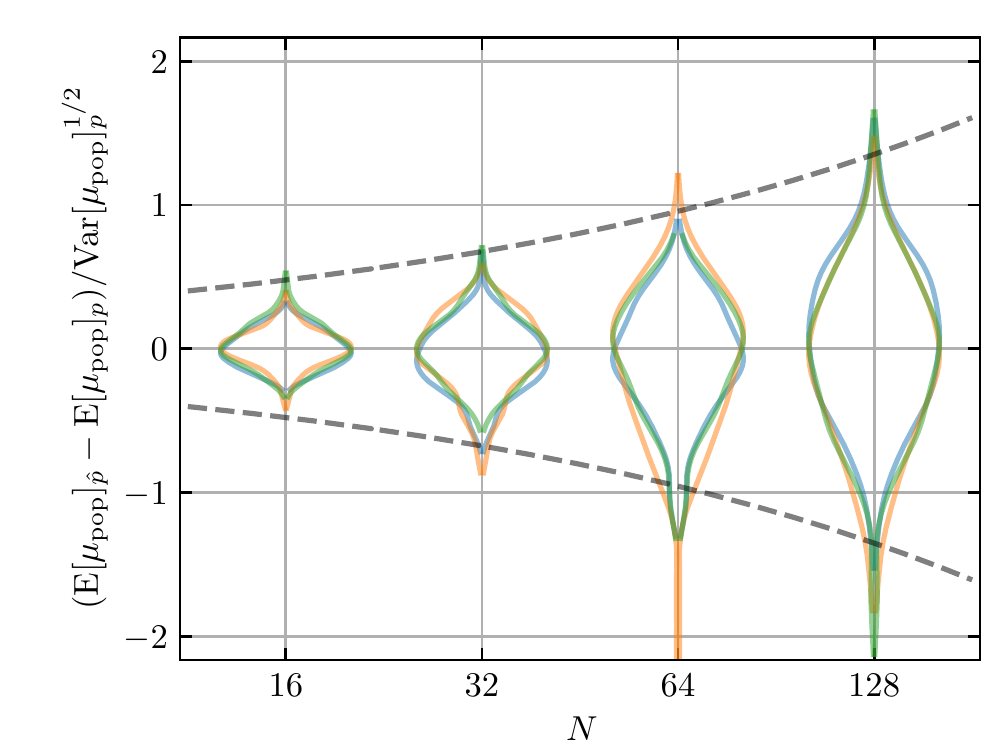}
        \end{center}
    \end{minipage}
    \begin{minipage}{0.49\textwidth}
        \begin{center}
            \large{$\quad \quad \quad m_i = 100$} \\
            \large{$\quad \quad \quad M = 100 N$} \\
            \includegraphics[width=1.0\textwidth]{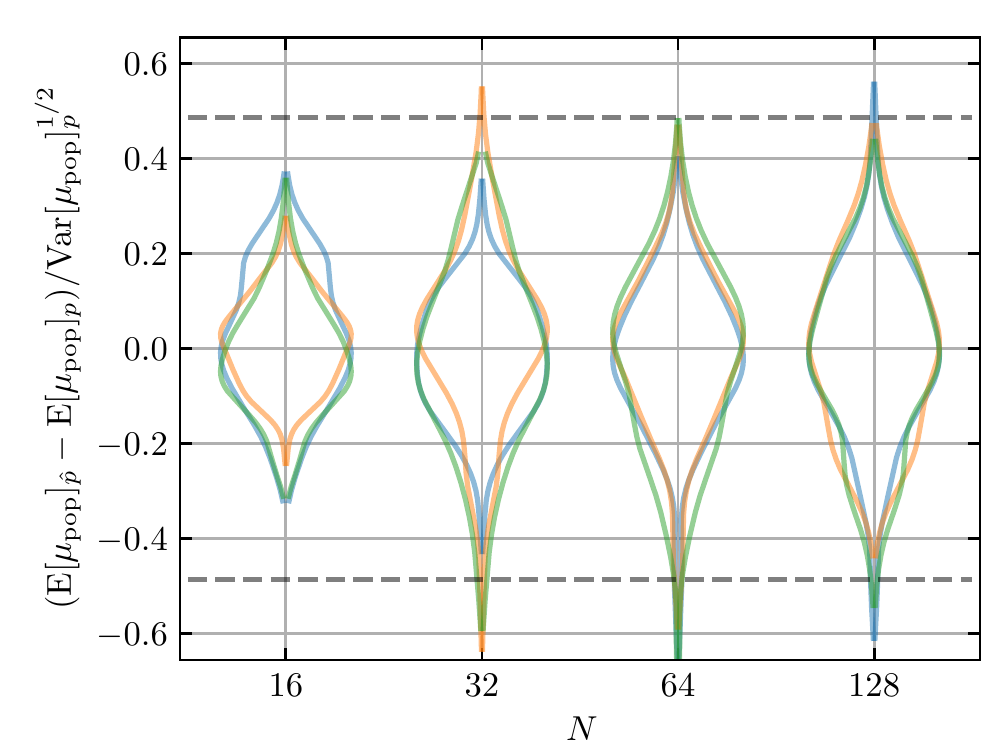}
        \end{center}
    \end{minipage}
    \caption{
        Analogous to Fig.~\ref{fig:everything-is-detectable} but for simulations that include selection effects.
        We consider a constant number of single-event posterior samples, with the number of injections (\emph{left}) fixed regardless of the catalog size or (\emph{right}) proportional to the catalog size.
        The number of injections must increase as the catalog grows, otherwise the Monte Carlo uncertainty can scatter the mean \textit{a posteriori} more than the nominal width of the posterior.
        In agreement with Eq.~\ref{eq:best criterion M scaling}, $M \propto N$ suffices to bound the relative error.
    }
    \label{fig:including-selection-effects}
\end{figure*}

Building upon our knowledge that keeping $m_i$ constant suffices, we investigate the impact of Monte Carlo uncertainty in estimates of the selection function.
We investigate two possible scalings, one in which the number of injections is constant\footnote{This scaling is what would be predicted by the criterion $\mathrm{Var}[\ln \hat{p}]_\mathrm{MC} \ll \mathrm{E}[\ln \hat{p}]^2_\mathrm{MC}$.} and one in which it scales linearly with the catalog size.
Fig.~\ref{fig:including-selection-effects} shows the result.

The size of the uncertainty from different Monte Carlo realizations can quickly become larger than the width of the posterior if the number of injections is constant as the catalog grows.
However, the relative uncertainty is constant if the number of injections is proportional to the catalog size.
The simulation shows that Eq.~\ref{eq:best criterion M scaling} is the correct scaling relation to perform precise population inference as inexpensively as possible.

As a final note, we remark that scaling $M$ more rapidly with the size of the catalog will not improve the precision of the inference unless the number of single-event posterior samples per event is also scaled with the size of the catalog.
Otherwise, additional injections will still reduce the uncertainty in the selection function, but that alone will not reduce the uncertainty in the total target distribution.
However, we have confirmed that the scaling $m_i \sim N$ and $M \sim N^2$ does reduce the relative uncertainty in the inference from the Monte Carlo sums as the catalog grows.


\subsubsection{Diagnosing Breakdown in the Estimator for the Target Distribution}
\label{sec:diagnosing breakdown}

Based on this model, we can identify phenomenological features of the target distribution when there are too few Monte Carlo samples.
In many cases, Monte Carlo variations produce $\hat{p}$ that resemble the $p$ that would be obtained with slightly different data, e.g., a slight shift in the maximum \textit{a posteriori}.
See Sec.~\ref{sec:generative modeling} for more discussion.
However, more severe issues may also occur.

Specifically, too few samples may introduce artificial maxima and minima into the target distribution (variance is large, so the estimate may be scattered widely).
If the Monte Carlo fluctuation produces a local maximum in the target distribution, then the inference will find support near that peak.
In this case, it is relatively straightforward to diagnose the issue.
Estimating the effective number of samples of an estimator, defined as $N_\mathrm{eff} = \mathrm{E}_\mathrm{MC}^2 / \mathrm{Var}_\mathrm{MC}$, throughout the posterior's support will identify situations in which the variance is large ($N_\mathrm{eff}$ is low).
Anecdotally, this may be most common for errors associated with $\hat{\mathcal{E}}$, not only because it enters the rate-marginalized target distribution with an exponent that depends on the catalog's size, but also because the rate-marginalized target distribution is inversely proportional to the selection function.
The selection function for many surveys will be relatively small (flux limited), and therefore Monte Carlo fluctuations may tend to scatter the estimate of the selection function to extremely small values.
These, then, produce very large peaks in the estimate of the target distribution.

This behavior can occur even if one samples from an estimate of the full hierarchical likelihood instead of the rate-marginalized likelihood.
This is because $p(\mathcal{R},\Lambda|\{D_i\}) = p(\mathcal{R}|\Lambda,\{D_i\}) p(\Lambda|\{D_i\})$ and $p(\mathcal{R}|\Lambda,\{D_i\}) = f(\mathcal{R}\mathcal{E}, N)$ when $p(\mathcal{R}) \sim \mathcal{R}^n$.
Errors in $\hat{\mathcal{E}}$ may make $p(\Lambda|\{D_i\})$ very peaked, and $\mathcal{E}$ simply sets the scale for $\mathcal{R}$.
That is, we can always choose an $\mathcal{R}$ for different $\mathcal{E}$ such that $p(\mathcal{R}|\Lambda,\{D_i\})$ is the same.
As such, when $\hat{\mathcal{E}}$ is scattered close to zero, $\hat{p}(\Lambda|\{D_i\})$ has a sharp peak in the neighborhood where $\hat{\mathcal{E}} \sim 0$.
The size and shape of this neighborhood is set by correlations in $\hat{\mathcal{E}}(\Lambda)$, not by the data.
This drives the maximum \textit{a posteriori} rate $\mathcal{R} \rightarrow \infty$ because $p(\mathcal{R}|\Lambda,\{D_i\})$ has a peak near $\mathcal{R}\mathcal{E} \sim N+n$ for $p(\mathcal{R}) \sim R^n$.
Thus, a very tightly peaked posterior for $\Lambda$ with divergent $\mathcal{R}$ often signals there are too few injections for precise inference.

The case where the Monte Carlo fluctuation produces a local minimum may be more difficult to diagnose \textit{post hoc} because samples drawn from the target distribution will naturally avoid such areas.
One may wish to draw samples from a tempered target distribution, and then estimate the effective number of samples throughout the tempered posterior.
This may allow one to more effectively explore local minima, at which point checking $N_\mathrm{eff}$ may also be more effective.

However, the best sanity check may be the analyst's experience.
If the shape of the posterior tends to favor or exclude models where $N_\mathrm{eff}$ is expected to be small, one will need to demonstrate that this is due to the data and not Monte Carlo fluctuations.
The best check is to increase $N_\mathrm{eff}$ for such models by drawing additional Monte Carlo samples and demonstrating that the target distribution is stable.
We return to these considerations in Sec.~\ref{sec:discussion}.


\section{Generative Model for Monte Carlo Samples}
\label{sec:generative modeling}

As we have seen, Monte Carlo uncertainty in the point estimate of the hierarchical population posterior can shift the posterior by more than the width of the true posterior.
We now consider how one might be able to model the uncertainty from the finite number of Monte Carlo samples self-consistently within the inference so that the width of the posterior would naturally account for the additional Monte Carlo uncertainty.
We do this by constructing a generative model for how Monte Carlo samples are obtained within the inference.

Let us begin by considering how a set of $m$ posterior samples is generated for an individual event along with the data and the true parameters for that event.
Consider a set of $m$ samples $\{\theta_j\}_{\theta|D,\Lambda_0}$ drawn from $p(\theta|D,\Lambda_0)$.
Then
\begin{align}
    p(\{\theta_j\}_{\theta|D,\Lambda_0}, D, \theta|\Lambda, \Lambda_0)
        & = \left[ \prod\limits_j^m p(\theta_j|D, \Lambda_0) \right] p(D|\theta) p(\theta|\Lambda)
\end{align}
This is just the usual model of how data is generated for an event with parameters $\theta$, which is in turn drawn from a population described by $\Lambda$, with the extra step of generating samples from $p(\theta|D,\Lambda_0)$.
In practice, we marginalize over the true signal parameters $\theta$ in our inference for $\Lambda$.
However, within this context, we also only use the Monte Carlo samples $\{\theta_j\}_{\theta|D,\Lambda_0}$; we do not use $D$ directly.
Therefore, we must also marginalize over $D$.
This yields a likelihood of obtaining the posterior samples given the population model (and a reference prior)
\begin{equation}
    p(\{\theta_j\}_{\theta|D,\Lambda_0}|\Lambda, \Lambda_0) = \int dD \left[ p(D|\Lambda) \prod\limits_j^m p(\theta_j|D, \Lambda_0) \right] \label{eq:super fancy single event}
\end{equation}
We can then condition the joint distribution of both $\{\theta_j\}_{\theta|D,\Lambda_0}$ and $\Lambda$ constructed from Eq.~\ref{eq:super fancy single event} and a prior for $\Lambda$ on the observed posterior samples in order to construct a posterior for $\Lambda$, just as one would normally condition on the data.
In this way, we see that $\{\theta_j\}_{\theta|D,\Lambda_0}$ serve as an imperfect (lossy) encoding of the data and detector sensitivity at the time of the event.
As $m \rightarrow \infty$, we expect the $\prod_j^m p(\theta_j|D,\Lambda_0)$ to be sharply peaked around the actual data that was recorded.
The additional uncertainty from Monte Carlo summation is associated with the information lost in this encoding.

Furthermore, we note that
\begin{align}
    \prod\limits_j^m p(\theta_j|D,\Lambda_0)
        & = \left[\prod\limits_j^m p(\theta_j|D_\mathrm{true},\Lambda_0)\right] \exp\left(-\sum\limits_j^m \ln \frac{p(\theta_j|D,\Lambda_0)}{p(\theta_j|D_\mathrm{true},\Lambda_0)} \right) \nonumber \\
        & \approx \left[\prod\limits_j^m p(\theta_j|D_\mathrm{true},\Lambda_0)\right] \exp\left(-m \int d\theta p(\theta|D_\mathrm{true},\Lambda_0) \ln \frac{p(\theta_j|D,\Lambda_0)}{p(\theta_j|D_\mathrm{true},\Lambda_0)} \right)
\end{align}
where $D_\mathrm{true}$ is the actual data so that $\theta_j \sim p(\theta|D_\mathrm{true},\Lambda_0)$ and we have approximated the sum over $\theta_j$ with an integral following the normal Monte Carlo rule.
In the last line, we recognize the Kullback-Leibler divergence
\begin{equation}
    D_\mathrm{KL}(D_\mathrm{true} || D) \equiv \int d\theta p(\theta|D_\mathrm{true},\Lambda_0) \ln \frac{p(\theta_j|D,\Lambda_0)}{p(\theta_j|D_\mathrm{true},\Lambda_0)}
\end{equation}
for the single-event posterior with respect to different observed data.
We note that $D_\mathrm{KL} \geq 0$ with equality only if $D=D_\mathrm{true}$.
Therefore, $p(\{\theta_j\}|\Lambda,\Lambda_0)$ is exponentially distributed in $D_\mathrm{KL}(D_\mathrm{true}||D)$ (Gaussian distributed in $(D-D_\mathrm{true})$ when this is small~\cite{10.5555/3019383}) with a scale parameter (inverse covariance) that scales with the number of Monte Carlo samples.
While it is readily apparent that larger $m$ imply more precise inference, it is difficult to state exactly how large
$m$ must be to meet practical considerations without including more details that are specific to the inference at hand.

In the context of our toy model, Eq.~\ref{eq:super fancy single event} becomes
\begin{equation}
    p(\{x_j\}|\Lambda) \propto \left[2\pi \left(\sigma^2_\mathrm{obs}\left(1+\frac{1}{m}\right) + \sigma^2_\mathrm{pop}\right)\right]^{-1/2} \exp\left( -\frac{\left(\mu_\mathrm{pop} - \frac{1}{m}\sum_j^m x_j\right)^2}{2\left(\sigma^2_\mathrm{obs}\left(1+\frac{1}{m}\right) + \sigma^2_\mathrm{pop}\right)}\right)
\end{equation}
\textit{De facto}, then, we see that Monte Carlo uncertainty increases the observational noise within the inference.
This matches our observation that different Monte Carlo realizations with the same catalog of detections produce estimators for the target distribution that appear similar to catalogs with slightly different detected data.

We also see that, as long as $m \gg 1$, the impact on the individual single-event evidence will be small.
This conclusion holds for $N$ events as well;\footnote{The demonstration of this is within the context of our toy model left as an exercise for the reader.} the impact of Monte Carlo uncertainty for individual single-event evidences does not scale with the size of the catalog, in agreement with Eq.~\ref{eq:best criterion m_i scaling}.

A similar model could be constructed for injections, the detector data, and detector sensitivity throughout the experiment.
We would need to marginalize over the detector sensitivity and data observed throughout the entire experiment.
However, as even the simpler integral over data for a single event in Eq.~\ref{eq:super fancy single event} is intractable in practice, such a model may be difficult to use.
We nonetheless note that, just as single-event posterior samples are a lossy encoding of the single-event likelihood, injections are a lossy encoding of the detector sensitivity throughout the experiment.


\section{Estimators Marginalized over Monte Carlo Uncertainty}
\label{sec:improved estimators}

Up until now, we have either considered the behavior of estimators for the target function constructed from point estimates of the Monte Carlo sums (Sec.~\ref{sec:general precision requirements}) or constructed generative models and marginalized over the observed data in the presence of lossy encoding from observed data to Monte Carlo samples and injections (Sec.~\ref{sec:generative modeling}).
As the generative models involve marginalization over the observed data, which is difficult to perform in practice, we now consider alternative approaches to account for Monte Carlo uncertainty.

Other authors have proposed marginalizing over the Monte Carlo uncertainty as a way to improve the behavior of the estimated target function (i.e., Ref.~\cite{Farr2019}).
We explore this idea further.
Specifically, we are interested in the expected value taken with respect to uncertainty in $\mathcal{Z}$ given the posterior samples and uncertainty in $\mathcal{E}$ given the injections.
\begin{equation}
    \mathrm{E}[\hat{p}]_{\begin{subarray} \mathcal{Z}_i|\mathrm{samp} \\ \mathcal{E}|\mathrm{inj}\end{subarray}} = \int \left[ \left(d\mathcal{E} p(\mathcal{E}|\mathrm{inj}) \prod\limits_i^N d\mathcal{Z}_i p(\mathcal{Z}_i|\mathrm{samp}) \right) \hat{p} \right]
\end{equation}
The marginalization factors so that each single-event evidence contributes a factor of
\begin{equation}
    \mathrm{E}[\mathcal{Z}_i]_{\mathcal{Z}_i|\mathrm{samp}} = \int d\mathcal{Z}_i p(\mathcal{Z}_i|\mathrm{samp}) \mathcal{Z}_i
\end{equation}
Appendix~\ref{sec:characteristic function} shows that $\mathrm{E}[\mathcal{Z}_i]_{\mathcal{Z}_i|\hat{\mathcal{Z}}_i,m_i} = \hat{\mathcal{Z}}_i$ with the exact Monte Carlo measure.
Therefore, the target function marginalized over Monte Carlo uncertainty in $\mathcal{Z}_i$ is identical to the point estimate.
Note that this marginalization, then, does not account for the variance from Monte Carlo uncertainty, unlike the generative models in Sec.~\ref{sec:generative modeling}.

Depending on whether we consider the full hierarchical likelihood or the rate-marginalized likelihood, we may be interested in either
\begin{equation}
    \mathrm{E}[e^{-\mathcal{R}\mathcal{E}}]_{\mathcal{E}|\mathrm{inj}} = \int d\mathcal{E} p(\mathcal{E}|\mathrm{inj}) e^{-\mathcal{R}\mathcal{E}}
\end{equation}
or
\begin{equation}
    \mathrm{E}[(\mathcal{E} + r^{-1})^{-(N+n)}]_{\mathcal{E}|\mathrm{inj}} = \int d\mathcal{E} \frac{p(\mathcal{E}|\mathrm{inj})}{(\mathcal{E} + r^{-1})^{N+n}}
\end{equation}
These marginalizations are not tractable with the full Monte Carlo measure.
Instead, Sec.~\ref{sec:approximate marginalization} considers several approximations to $p(\mathcal{E}|\mathrm{inj})$, and we discuss their behavior in relation to the point estimate in Sec.~\ref{sec:comparison of estimators}.


\subsection{Marginalization with Approximate Models of Monte Carlo Uncertainty in the Selection Function}
\label{sec:approximate marginalization}

We investigate several approximate measures of Monte Carlo uncertainty and explore whether marginalized estimators derived therefrom provide better estimates for the target distribution.
That is, we assume a functional form for the uncertainty in the true selection function given the Monte Carlo samples\footnote{This is usually done by matching the moments of the model to estimates of the moments of the Monte Carlo estimator under different realizations of the sample set.} and then perform the marginalization analytically.
If we are concerned with the full target distribution, then the relevant expectation value is $\mathrm{E}[e^{-\mathcal{R}\mathcal{E}}]_{\mathcal{E}|\mathrm{inj}}$.
If we are instead concerned with the rate-marginalized target distribution assuming a $\Gamma$-distribution prior for $\mathcal{R}$, we are concerned with $\mathrm{E}[(\mathcal{E} + r^{-1})^{-(N+n)}]_{\mathcal{E}|\mathrm{inj}}$.
At times, we will limit ourselves to the limit $r\rightarrow\infty$ so that $p(\mathcal{R}) \sim \mathcal{R}^{n-1}$.

We consider two simple models for the Monte Carlo uncertainty: Gaussian uncertainty (Sec.~\ref{sec:gaussian}) and log-normal uncertainty (Sec.~\ref{sec:lognormal}).
Sec.~\ref{sec:comparison of estimators} compares the behavior of these marginalized estimators to the point estimate.


\subsubsection{Gaussian Model for $p(\mathcal{E}|\mathrm{inj})$}
\label{sec:gaussian}

As a first approximation, we model the uncertainty in $\mathcal{E}$ as a Gaussian
\begin{equation}
    p(\mathcal{E}|\mathrm{inj}) \sim \frac{1}{\sqrt{2\pi\sigma^2}} e^{-(\mathcal{E} - \hat{\mathcal{E}})^2 / 2\hat{\sigma}^2}
\end{equation}
where
\begin{equation}
    \hat{\sigma}^2 = \frac{1}{m}\left[\frac{1}{m-1}\sum\limits_k^m \left(\frac{mp(\theta_k|\Lambda)}{Mp(\theta_k|\mathrm{draw})} - \hat{\mathcal{E}}\right)^2\right]
\end{equation}
approximates the variance of the Monte Carlo sum from the sample variance of the Monte Carlo sample set.
This has already been considered in detail elsewhere~\cite{Farr2019}, and it was noted that this model leads to bad behavior because there is support for $\mathcal{E} \leq 0$.
Specifically,
\begin{equation}
    \mathrm{E}[e^{-\mathcal{R}\mathcal{E}}]_{\mathcal{E}|\mathrm{inj}} = e^{-\mathcal{R}\hat{\mathcal{E}} + \frac{1}{2}\mathcal{R}^2\hat{\sigma}^2} \label{eq:gaussian exp -re}
\end{equation}
which diverges as $\mathcal{R}\rightarrow\infty$.
Ref.~\cite{Farr2019} notes that the product $\mathcal{R}^N e^{-\mathcal{R}\hat{\mathcal{E}} + \hat{\sigma}^2\mathcal{R}^2/2}$ has a local maximum iff $\hat{\mathcal{E}}^2 > 4 N \hat{\sigma}^2$.
This motivates the requirement\footnote{
The scaling required for the Gaussian approximation to hold just happens to be the same as we found in Sec.~\ref{sec:general precision requirements}.
This is entirely due to the assumed functional form for $p(\mathcal{E}|\mathrm{inj})$.
Other assumptions do not impose such a requirement.
} that $\hat{\mathcal{E}}^2/\hat{\sigma}^2 \equiv N_\mathrm{eff} > 4 N$.
Ref.~\cite{Farr2019} then argues that marginalization over $\mathcal{R}$, which formally always diverges, may be approximated by a Gaussian centered on the local maximum with an approximation of its width.
This yields
\begin{equation}
    \frac{\mathrm{E}[\mathcal{E}^{-N}]_{\mathcal{E}|\mathrm{inj}}}{\mathrm{E}[\mathcal{E}]_{\mathcal{E}|\mathrm{inj}}^{-N}} = \exp\left( \frac{3N + N^2}{2N_\mathrm{eff}} \right)
\end{equation}
which we explore alongside other approximations to the Monte Carlo uncertainty in Sec.~\ref{sec:comparison of estimators}.
Briefly, this requires $N_\mathrm{eff} > \mathcal{O}(N^2)$ for the marginalized estimate to collapse to the point estimate at any individual point in parameter space.
This is the same scaling as Eq.~\ref{eq:sufficient M scaling}.
In order for the point estimate to provide precise inference at any individual point in parameter space, we required $N_\mathrm{eff} \gg N^2$.

For completeness, we note that
\begin{equation}
    \mathrm{E}[(\mathcal{E} + r^{-1})^{-(N+n)}]_{\mathcal{E}|\mathrm{inj}} = \frac{1}{\sqrt{2\pi\sigma^2}} \int d\mathcal{E} \frac{e^{-(\mathcal{E}-\hat{\mathcal{E}})^2/2\hat{\sigma}^2}}{(\mathcal{E} + r^{-1})^{N+n}}
\end{equation}
always diverges because we integrate through $\mathcal{E} = -r^{-1}$.
The divergence in Eq.~\ref{eq:gaussian exp -re} overwhelms the Gamma-distribution prior $\sim e^{-\mathcal{R}/r}$ as $\mathcal{R} \rightarrow \infty$.
We therefore conclude that a Gaussian model for the Monte Carlo uncertainty in the selection function may never suffice, and it may break down particularly badly in any part of parameter space for which $N_\mathrm{eff} \lesssim 4 N$.
That is, unless $N_\mathrm{eff} > 4N$, stochastic samples will never become trapped in the local maximum in $\hat{p}$ and instead will always explore the tail as $\mathcal{R} \rightarrow \infty$.


\subsubsection{Log-Normal Model for $p(\mathcal{E}|\mathrm{inj})$}
\label{sec:lognormal}

Motivated by the fact that the Gaussian model breaks down, in part, because it has support where $\mathcal{E} \leq 0$, we consider a model that automatically excludes this region
\begin{equation}
    p(\mathcal{E}|\mathrm{inj}) = \frac{\mathcal{E}^{-1}}{\sqrt{2\pi\sigma^2}} e^{-(\ln \mathcal{E} - \mu)^2/2\sigma^2}
\end{equation}
While $\mathrm{E}[e^{-\mathcal{R}\mathcal{E}}]_{\mathcal{E}|\mathrm{inj}}$ is not easily expressed in closed form, we note that the integral always exists.
Similarly, the expectation value relevant for the rate-marginalized likelihood with a $\Gamma$ prior on $\mathcal{R}$ is difficult to express in closed form in general, but in the limit $r \rightarrow \infty$ we find
\begin{align}
    \mathrm{E}[\mathcal{E}^{-N}]_{\mathcal{E}|\mathrm{inj}}
        & = e^{-N\mu + \frac{1}{2}\sigma^2N^2} \nonumber \\
        & = \mathrm{E}[\mathcal{E}]_{\mathcal{E}|\mathrm{inj}}^{-N} e^{\frac{1}{2}\sigma^2 N(N+1)}
\end{align}
This expectation value also always exists and, therefore, there is no requirement on $N_\mathrm{eff}$ for the approximation to hold.
Again, the effective number of samples is $N_\mathrm{eff} \equiv \mathrm{E}[\mathcal{E}]_{\mathcal{E}|\mathrm{inj}}^2 / \mathrm{Var}[\mathcal{E}]_{\mathcal{E}|\mathrm{inj}} = (e^{\sigma^2} - 1)^{-1}$, which implies $\sigma^2 = \ln(1+1/N_\mathrm{eff})$.
In the limit $N_\mathrm{eff} \gg 1$,
\begin{equation}
    \frac{\mathrm{E}[\mathcal{E}^{-N}]_{\mathcal{E}|\mathrm{inj}}}{\mathrm{E}[\mathcal{E}]_{\mathcal{E}|\mathrm{inj}}^{-N}} \approx \exp\left( \frac{N(N+1)}{2N_\mathrm{eff}} \right)
\end{equation}
We again need $N_\mathrm{eff} \sim \mathcal{O}(N^2)$ in order for the point estimate to be a good estimate for the marginalized target distribution for this approximation as well.
That is, even though the marginalization will always converge, we need the number of injections to scale quadratically with the catalog size in order for the point-estimate of the target distribution to be a reliable estimator for the target distribution at any individual point in parameter space.

Finally, we note that $\mathrm{E}[\mathcal{E}^{-N}]_{\mathcal{E}|\mathrm{inj}}$ can be larger than one when $\sigma^2 N^2$ is not small.
This is outside the allowed domain ($0 \leq \mathcal{E}(\Lambda) \equiv P(\mathrm{det}|\Lambda) \leq 1$), and this is because the log-normal model has support for $\mathcal{E} > 1$.\footnote{
$\hat{\mathcal{E}}$ may be $> 1$ even though $\mathcal{E}$ cannot be.
}
This tail will dominate the marginalization if the Monte Carlo uncertainty is large.
One may alternatively consider a truncated log-normal model which includes a Heaviside function to enforce the requirement that $\mathcal{E} \leq 1$.
Marginalization still yields a correction that roughly scales as $\mathcal{O}(e^{N^2/N_\mathrm{eff}})$.
Other models, such as a $\beta$-distribution, also produce corrections that scale in the same way.
As this scaling is the primary consideration in our comparison to the point estimate, we do not pursue additional models in greater detail.


\subsubsection{Comparison of Estimators}
\label{sec:comparison of estimators}

We typically find $\mathrm{E}[\mathcal{E}^{-N}]_{\mathcal{E}|\mathrm{inj}} \sim \hat{\mathcal{E}}^{-N} e^{N^2/N_\mathrm{eff}}$.
The effective change in the maximum of the target distribution is then determined by a derivative of the extra contribution.
That is, the maximum likelihood is where
\begin{equation}
    \partial_\Lambda \ln \mathrm{E}[\mathcal{E}^{-N}]_{\mathcal{E}|\mathrm{inj}} \sim -N \left( \partial_\Lambda \ln \hat{\mathcal{E}} + \frac{N}{N_\mathrm{eff}} \partial_\Lambda \ln N_\mathrm{eff} \right) = 0
\end{equation}
Furthermore, as Ref.~\cite{Farr2019} points out, it can be shown that $\partial_\mu \ln N_\mathrm{eff}$ does not depend on the number of Monte Carlo samples, and so the shift introduced by the modification to the likelihood scales as $\mathcal{O}(N/N_\mathrm{eff})$.
Additionally, the size of the posterior support still collapses as $\mathrm{Cov}[\Lambda_a, \Lambda_b]_{p(\Lambda|\{D_i\})} \sim 1/N$.
This means that, for large $N$, the shift will eventually be larger than the posterior uncertainty if $N_\mathrm{eff} \sim N$.
This means the marginalized estimate will collapse to a different $\Lambda$ than the point estimate, and we know that the point estimate collapses to the true population parameters in this limit.
In order to guarantee that the shift shrinks faster than the posterior uncertainty, we therefore require $N_\mathrm{eff} \gtrsim N^{3/2}$.
As such, marginalization saved us from the $\mathcal{O}(N^2)$ scaling in Eq.~\ref{eq:sufficient M scaling}, but it is still more expensive than using the point estimate directly.

This is because the marginalization, as conducted in this Section, treats the estimate of the target function at each point in parameter space as statistically independent.
That is not the case, and we have seen that correlations between the estimator at neighboring points in parameter space are crucial for the precision of the inference. 
The inference with the marginalized target distribution, then, requires additional Monte Carlo samples compared to the point estimate to resolve the target distribution well enough to overcome the bias introduced by this approximation.

We note that one may instead try to marginalize over the Monte Carlo uncertainty in the target function at all points in parameter space simultaneously to incorporate such correlations.
However, to do so, one would need to compute the point estimate at every point in parameter space at the same time.
Pragmatically, this would mean evaluating many Monte Carlo sums on a grid throughout a high dimensional parameter space, which will be exponentially costly in the dimensionality of the model.
Therefore, for all practical purposes, one is likely best served by using the point estimate for the target distribution.


\section{Discussion}
\label{sec:discussion}

We explored precision requirements for Monte Carlo sums used to estimate the target distribution within hierarchical Bayesian inference.
We showed that properly accounting for correlations between estimators evaluated at neighboring points in parameter space with the same set of Monte Carlo samples, along with the reduction in size of the posterior uncertainty as the size of the catalog grows, produces the following requirements.
\begin{itemize}
    \item The number of posterior samples for each individual event required for precise inference is independent of the catalog size.
    \item The number of injections required to estimate the selection function precisely enough scales linearly with the catalog size.
\end{itemize}
These scalings were verified in the context of an analytically tractable toy model.
We also showed that the additional uncertainty from Monte Carlo sample sets arises because they are imperfect representations of the data recorded for each event and the survey sensitivity throughout the experiment by considering a probabilistic model of how Monte Carlo samples are generated.

Furthermore, we investigated the behavior of target distributions that were marginalized over approximations of the Monte Carlo uncertainty in the selection function.
We find that such estimators may be biased more than the size of the posterior uncertainty in the limit of many Monte Carlo samples unless the number of injections scales faster than the size of the catalog.
This makes precise inference with marginalized estimators more expensive than using the point estimate, which we attribute to the fact that the marginalization neglects correlations between the Monte Carlo estimate at neighboring points in parameter space.
Attempts to compute the joint distribution of the selection function at different point in parameter space simultaneously from Monte Carlo estimates evaluated throughout parameter space are likely to suffer from the curse of dimensionality and are likely impractical computationally.

Pragmatically, then, we suggest the following.
Analysts should always use target distributions constructed from Monte Carlo point estimates.
They may proceed in one of two ways.

Analysts may compute the variance under Monte Carlo uncertainty between combinations of any two sets of population parameters drawn from the prior.
This can be done with estimates of the variance of Monte Carlo sums obtained from the variance within the Monte Carlo sample set.
See Appendix~\ref{sec:estimators}.
If the variance of the target distribution is small everywhere in the prior, then there are sufficient samples to perform accurate inference.
However, this will likely require $\mathcal{O}(N)$ single-event posterior samples per event and $\mathcal{O}(N^2)$ injections.

Alternatively, analysts may initially draw what they believe to be a reasonable number of samples and injections and then sample from the estimate of the target distribution.
They can then estimate the variance in $\Delta \ln \hat{p}$ between pairs of points drawn from the posterior (rather than the prior).
If the variance of $\Delta \ln \hat{p}$ between any two points drawn from the posterior is large, then more samples must be drawn and the posterior must be recalculated.
This procedure should be iterated until the variance is small throughout the posterior.
Although the process of iterating may be expensive, we expect the number of Monte Carlo samples required will scale better with the catalog size.
What's more, it is straightforward to determine which term in $\Delta \ln \hat{p}$ dominates the Monte Carlo uncertainty.
This allows one to draw additional samples only where needed, e.g., for only some single-event evidences or only additional injections in certain parts of parameter space.

However, we also note that precision requirements set by expressions like
\begin{equation}
    \mathrm{Var}[\hat{X}(\Lambda)]_\mathrm{MC} = \frac{f(\Lambda)}{m} \label{eq:this one}
\end{equation}
may indirectly depend on the catalog size because analysts may chose to examine different $\Lambda$ depending on the catalog.
For example, more detected events may encourage analysts to look for narrow or sharp features within their population model.
Such features may have intrinsically larger Monte Carlo variance (larger $f(\Lambda)$ in Eq.~\ref{eq:this one}) and therefore require larger $m$ to perform precise inference.
As such, analysts' choices of what parts of parameter space to explore may introduce additional problem-specific scalings between the number of Monte Carlo samples required and the catalog size.
These are difficult to diagnose in general, but are worth bearing in mind for any inference involving population models with sharp or narrow features.

Finally, we note that Monte Carlo estimates of high-dimensional integrals are known to converge slowly ($\mathrm{Var}_\mathrm{MC} \sim 1/N$).
Other sampling techniques, such as low-discrepancy sequences~\cite{NIEDERREITER198851}, may be able to speed-up the integral's convergence so that $\mathrm{Var}_\mathrm{MC} \sim (\ln N)^{2(N_\mathrm{dim}-1)}/N^2$.
Such quasi-Monte Carlo sampling procedures may provide faster convergence than normal Monte Carlo sampling for large $N$ and low dimensional dimensional integrals (small $N_\mathrm{dim}$).
These may be difficult to implement for single-event evidences,\footnote{
Even if low-discrepancy sequences were implemented for single-event evidences, they would not change the fact that the number of samples needed does not scale with the size of the catalog.
They would simply reduce the number of samples needed without altering the scaling with the catalog size.
}
but the current process of generating injections for estimates of the selection function may be more amenable.
That is, if the injection distribution is completely separable so that
\begin{equation}
    p(\theta_1, \theta_2, \cdots, \theta_{N_\mathrm{dim}}|\mathrm{draw}) = \prod\limits_i^{N_\mathrm{dim}} p_i(\theta_i|\mathrm{draw})
\end{equation}
then we may be able to approximate high-dimensional integrals with respect to this measure with a low-discrepancy sequence.
If so, then the number of injections required for precise inference may only scale as $\mathcal{O}(N^{1/2})$.
This may provide significant computational savings for catalogs with $\gtrsim O(10^2-10^3)$ events expected from advanced and third generation (3G) GW detectors operating at design sensitivity~\cite{Abbott2018, Abbott2017, Evans2021}.
Low-discrepancy sequences are not without their own complications, though, and additional study may be needed before they can be adopted for empirical estimates of the selection function within hierarchical inference over GW catalogs.


\acknowledgments

The authors are extremely grateful to Maya Fishbach for many helpful discussions throughout the preparation of this manuscript.
Research at Perimeter Institute is supported in part by the Government of Canada through the Department of Innovation, Science and Economic Development Canada and by the Province of Ontario through the Ministry of Colleges and Universities.
R.E. also thanks the Canadian Institute for Advanced Research (CIFAR) for support.
The  Flatiron  Institute  is  supported  by  the  Simons Foundation.


\bibliography{biblio}


\appendix


\section{Moments of Monte Carlo Estimators}
\label{sec:estimators}

This appendix provides expressions for the moments of Monte Carlo estimators.
Sec.~\ref{sec:single-event evidences} examines $\hat{\mathcal{Z}}$.
Sec.~\ref{sec:selection function} examines $\hat{\mathcal{E}}$.
Finally, Sec.~\ref{sec:metric} examines the variance of the difference of the log of our estimator for the target function at two nearby points.


\subsection{Single-Event Evidences}
\label{sec:single-event evidences}

Because we control the number of samples drawn from each event's reference posterior, the measure for Monte Carlo approximations to the single-event evidence is straightforward.
The expectation value is
\begin{align}
    \mathrm{E}[\hat{\mathcal{Z}}_i]_\mathrm{MC}
        & = \int \left[\prod\limits_j^{m_i} d\theta_j\, p(\theta_j|D_i,\Lambda_0) \frac{p(D_i|\Lambda_0)}{m_i}\sum_k^{m_i}\frac{p(\theta_k|\Lambda)}{p(\theta_k|\Lambda_0)} \right] \nonumber \\
        & = p(D_i|\Lambda_0) \int d\theta\, p(\theta|D_i,\Lambda_0) \frac{p(\theta|\Lambda)}{p(\theta|\Lambda_0)} \nonumber \\
        & = \int d\theta\, p(\theta|\Lambda) p(D_i|\theta) \nonumber \\
        & = \mathcal{Z}_i \label{eq:Z expected value}
\end{align}
because each Monte Carlo sample is i.i.d..
We note that approximating the integral over single-event parameters with a Monte Carlo sum implies that Eq.~\ref{eq:Zhat} is an estimator for Eq.~\ref{eq:Z expected value} modulo a multiplicative constant that will not affect the inference.
Furthermore, the second moment is
\begin{align}
    \mathrm{E}[\hat{\mathcal{Z}}_i(\Lambda_a) \hat{\mathcal{Z}}_i(\Lambda_b)]_\mathrm{MC}
        & = \int \left[\prod\limits_j^{m_i} d\theta_j\, p(\theta_j|D_i,\Lambda_0)\right] \left[\frac{p(D_i|\Lambda_0)}{m_i}\sum_k^{m_i}\frac{p(\theta_k|\Lambda_a)}{p(\theta_k|\Lambda_0)} \right] \left[\frac{p(D_i|\Lambda_0)}{m_i}\sum_l^{m_i}\frac{p(\theta_l|\Lambda_b)}{p(\theta_l|\Lambda_0)} \right] \nonumber \\
        & = \int \left[\prod\limits_j^{m_i} d\theta_j\, p(\theta_j|D_i,\Lambda_0)\right] \frac{p(D_i|\Lambda_0)^2}{m_i^2}\left[ \sum\limits_k^{m_i} \left(\frac{p(\theta_k|\Lambda_a)}{p(\theta_k|\Lambda_0)}\right)\left(\frac{p(\theta_k|\Lambda_b)}{p(\theta_k|\Lambda_0)}\right) + \sum\limits_k^{m_i} \frac{p(\theta_k|\Lambda_a)}{p(\theta_k|\Lambda_0)} \sum\limits_{l\neq k}^{m_i} \frac{p(\theta_l|\Lambda_b)}{p(\theta_l|\Lambda_0)} \right] \nonumber \\
        & = \frac{p(D_i|\Lambda_0)^2}{m_i} \left[ \int d\theta\, p(\theta|D_i,\Lambda_0) \left(\frac{p(\theta|\Lambda_a)}{p(\theta|\Lambda_0)}\right)\left(\frac{p(\theta|\Lambda_b)}{p(\theta|\Lambda_0)}\right) \right. \nonumber \\
        & \quad\quad\quad\quad\quad\quad\quad\quad \left. + (m_i-1)\left(\int d\theta\, p(\theta|D_i,\Lambda_0) \frac{p(\theta|\Lambda_a)}{p(\theta|\Lambda_0)} \right)\left(\int d\theta\, p(\theta|D_i,\Lambda_0) \frac{p(\theta|\Lambda_b)}{p(\theta|\Lambda_0)} \right) \right]
\end{align}
so that the covariance is
\begin{equation}
    \mathrm{Cov}[\hat{\mathcal{Z}}_i(\Lambda_a), \hat{\mathcal{Z}}_i(\Lambda_b)]_\mathrm{MC} = \frac{p(D_i|\Lambda_0)^2}{m_i} \int d\theta\, p(\theta|D_i,\Lambda_0)\left( \frac{p(\theta|\Lambda_a)p(\theta|\Lambda_b)}{p(\theta|\Lambda_0)^2} - \frac{E[\hat{\mathcal{Z}}_i(\Lambda_a)] E[\hat{\mathcal{Z}_i}(\Lambda_b)]}{p(D_i|\Lambda_0)^2}\right) \label{eq:Z cov}
\end{equation}
with associated estimator
\begin{equation} \label{eq:cov Zhat}
    \widehat{\mathrm{Cov}}[\hat{\mathcal{Z}}_i(\Lambda_a), \hat{\mathcal{Z}}_i(\Lambda_b)]_\mathrm{MC} = \frac{p(D_i|\Lambda_0)^2}{m_i} \left[ \frac{1}{m_i-1} \sum\limits_j^{m_i} \left(\frac{p(\theta_j|\Lambda_a)}{p(\theta|\Lambda_0)} - \frac{\hat{\mathcal{Z}}_i(\Lambda_a)}{p(D_i|\Lambda_0)}\right) \left(\frac{p(\theta_j|\Lambda_b)}{p(\theta_j|\Lambda_0)} - \frac{\hat{\mathcal{Z}}_i(\Lambda_b)}{p(D_i|\Lambda_0)} \right) \right]
\end{equation}
where we have used the sample covariance and Eq.~\ref{eq:Zhat} to estimate the integrals over $\theta$ from the Monte Carlo sample set.
We note that the covariance between $\hat{\mathcal{Z}}$ evaluated at different population parameters will not vanish in general.
This is because $\hat{\mathcal{Z}}(\Lambda_a)$ and $\hat{\mathcal{Z}}(\Lambda_b)$ are computed with the same set of samples.


\subsection{Selection Function}
\label{sec:selection function}

The estimator's measure for the selection function is more complicated than for the single-event evidences because, while we control the total number of injections drawn ($M$), we only record the parameters of a subset of $m$ detected injections.
Therefore, the size of that set is subject to further sampling uncertainty.
The mean of our estimator is
\begin{align}
    \mathrm{E}[\hat{\mathcal{E}}]_\mathrm{MC}
        & = \sum\limits_{m=0}^{M} p(m|M) \int \left[ \left(\prod\limits_j^m d\theta_j p(\theta_j|\mathrm{det},\mathrm{draw})\right) \left(\frac{1}{M}\sum\limits_k^m \frac{p(\theta_k|\Lambda)}{p(\theta_k|\mathrm{draw})} \right)\right] \nonumber \\
        & = \sum\limits_{m=0}^M p(m|M) \left[ \frac{m}{MP(\mathrm{det}|\mathrm{draw})} \int d\theta\, p(\theta|\Lambda) P(\mathrm{det}|\theta) \right] \nonumber \\
        & = \left( \int d\theta\, p(\theta|\Lambda) P(\mathrm{det}|\theta) \right) \left( \frac{1}{MP(\mathrm{det}|\mathrm{draw})}\right) \sum\limits_{m=0}^{M} p(m|M) m \nonumber \\
        & = \int d\theta\, p(\theta|\Lambda) P(\mathrm{det}|\theta) \nonumber \\
        & = \mathcal{E}(\Lambda) \label{eq:E expected value}
\end{align}
because the number of detected events follows a binomial distribution
\begin{equation}
    p(m|M) = \left(\begin{matrix} M \\ m\end{matrix}\right) P(\mathrm{det}|\mathrm{draw})^m (1 - P(\mathrm{det}|\mathrm{draw}))^{M-m}
\end{equation}
with expected value $\mathrm{E}[m]_\mathrm{MC} = \sum_{0}^{M} m p(m|M) = MP(\mathrm{det}|\mathrm{draw})$.
The second moment is
\begin{align}
    \mathrm{E}[\hat{\mathcal{E}}(\Lambda_a)\hat{\mathcal{E}}(\Lambda_b)]_\mathrm{MC}
        & = \sum\limits_{m=0}^M p(m|M) \int \prod\limits_j^m d\theta_j\,p(\theta_j|\mathrm{det},\mathrm{draw}) \left(\frac{1}{M}\sum\limits_k^m \frac{p(\theta_k|\Lambda_a)}{p(\theta_k|\mathrm{draw})}\right) \left(\frac{1}{M}\sum\limits_l^m \frac{p(\theta_l|\Lambda_b)}{p(\theta_l|\mathrm{draw})}\right) \nonumber \\
        & = \frac{P(\mathrm{det}|\mathrm{draw})}{M} \int d\theta\, p(\theta|\mathrm{det},\mathrm{draw}) \left(\frac{p(\theta|\Lambda_a)}{p(\theta|\mathrm{draw})} \right) \left(\frac{p(\theta|\Lambda_b)}{p(\theta|\mathrm{draw})} \right) \nonumber \\
        & \quad\quad + \frac{P(\mathrm{det}|\mathrm{draw})^2 (M-1)}{M}\left( \int d\theta\, p(\theta|\mathrm{det},\mathrm{draw}) \frac{p(\theta|\Lambda_a)}{p(\theta|\mathrm{draw})}\right) \left( \int d\theta\, p(\theta|\mathrm{det},\mathrm{draw}) \frac{p(\theta|\Lambda_b)}{p(\theta|\mathrm{draw})}\right)
\end{align}
implying the covariance is
\begin{multline}
    \mathrm{Cov}[\hat{\mathcal{E}}(\Lambda_a), \hat{\mathcal{E}}(\Lambda_b)]_\mathrm{MC} \\
        = \frac{P(\mathrm{det}|\mathrm{draw})}{M}\left[ \int d\theta\, p(\theta|\mathrm{det},\mathrm{draw})\frac{p(\theta|\Lambda_a)p(\theta|\Lambda_b)}{p(\theta|\mathrm{draw})^2} - \left(\frac{E[\hat{\mathcal{E}}(\Lambda_a)] E[\hat{\mathcal{E}}(\Lambda_b)]}{P(\mathrm{det}|\mathrm{draw})^2}\right)
 \right] \\
        + \frac{P(\mathrm{det}|\mathrm{draw})(1-P(\mathrm{det}|\mathrm{draw}))}{M}\left(\frac{E[\hat{\mathcal{E}}(\Lambda_a)] E[\hat{\mathcal{E}}(\Lambda_b)]}{P(\mathrm{det}|\mathrm{draw})^2}\right) \label{eq:E cov}
\end{multline}
Appealingly, each term has a clear interpretation: the first term captures the variance between different $\theta_k$ in the found injections while the second captures the uncertainty due to the number of found injections.
We define an estimator for the covariance as follows
\begin{multline} \label{eq:cov Ehat}
    \widehat{\mathrm{Cov}}[\hat{\mathcal{E}}(\Lambda_a) \hat{\mathcal{E}}(\Lambda_b)]_\mathrm{MC} = \frac{1}{M} \left[ \frac{m}{(m-1)M}\sum\limits_k^m \left(\frac{p(\theta_k|\Lambda_a)}{p(\theta_k|\mathrm{draw})} - \frac{M}{m}\hat{\mathcal{E}}(\Lambda_a) \right) \left(\frac{p(\theta_k|\Lambda_b)}{p(\theta_k|\mathrm{draw})} - \frac{M}{m}\hat{\mathcal{E}}(\Lambda_b) \right) \right. \\ + \left. \frac{(M-m)}{m} \hat{\mathcal{E}}(\Lambda_a)\hat{\mathcal{E}}(\Lambda_b) \right]
\end{multline}
again using the sample variance to approximate integrals over $\theta$.


\subsection{$\mathrm{Var}[\Delta \ln \hat{p}]$ in Local Neighborhoods}
\label{sec:metric}

We begin by noting that the prior will not contribute to the Monte Carlo variance because it is known analytically.
We therefore focus on the variance of the estimate of the rate-marginalized hierarchical likelihood: $\Delta \ln \hat{p}(\{D_i\}|\Lambda)$, which can be decomposed into independent parts so that
\begin{align}
    \mathrm{Var}[\ln \hat{p}(\{D_i\}|\Lambda_a) - \ln \hat{p}(\{D_i\}|\Lambda_b)]_\mathrm{MC}
        & = (N+n)^2 \mathrm{Var}\left[\ln \left(\hat{\mathcal{E}}(\Lambda_a) + r^{-1}\right) - \ln \left(\hat{\mathcal{E}}(\Lambda_b) + r^{-1}\right)\right]_\mathrm{MC} \nonumber \\
        & \quad + \sum\limits_i^N \mathrm{Var}[\ln \hat{\mathcal{Z}}(\Lambda_a) - \ln \hat{\mathcal{Z}}(\Lambda_b)]_\mathrm{MC}
\end{align}
We then consider each of these terms in turn when $\Delta \equiv \Lambda_b - \Lambda_a$ is small.

Starting from Eqs.~\ref{eq:var delta lnz},~\ref{eq:Z expected value}, and~\ref{eq:Z cov}, we see that
\begin{align}
    \mathrm{Var}[\ln \hat{\mathcal{Z}}_i(\Lambda_a) - \ln \hat{\mathcal{Z}}_i(\Lambda_b)]_\mathrm{MC} 
        & = \frac{m_i^{-1}}{\mathcal{Z}(\Lambda_a)^2} \int d\theta p(\theta|D_i,\Lambda_0) \frac{p(\theta|\Lambda_a)^2}{p(\theta|\Lambda_0)^2}
          + \frac{m_i^{-1}}{\mathcal{Z}(\Lambda_b)^2} \int d\theta p(\theta|D_i,\Lambda_0) \frac{p(\theta|\Lambda_b)^2}{p(\theta|\Lambda_0)^2} \nonumber \\
        & \quad - \frac{2m_i^{-1}}{\mathcal{Z}(\Lambda_a)\mathcal{Z}(\Lambda_b)} \int d\theta p(\theta|D_i,\Lambda_0) \frac{p(\theta|\Lambda_a)p(\theta|\Lambda_b)}{p(\theta|\Lambda_0)^2}
\end{align}
Expanding the population prior around $\Lambda_a$ so that
\begin{equation}
    p(\theta|\Lambda_b) \approx p(\theta|\Lambda_a) + \left. \partial_\mu p(\theta|\Lambda)\right|_{\Lambda_a} \Delta^\mu \\
\end{equation}
yields
\begin{align}
    \mathrm{Var}\left[ \ln \hat{\mathcal{Z}}_i(\Lambda_a) \right. & \left. - \ln \hat{\mathcal{Z}}_i(\Lambda_b)\right]_\mathrm{MC} = \nonumber \\
        & \frac{m_i^{-1}}{\mathcal{Z}(\Lambda_a)^2} \int d\theta p(\theta|D_i,\Lambda_0) \frac{p(\theta|\Lambda_a)^2}{p(\theta|\Lambda_0)^2} \nonumber \\
        & + \frac{m_i^{-1}}{\mathcal{Z}(\Lambda_a)^2} \left( 1 - 2\frac{\partial_\mu \mathcal{Z}|_{\Lambda_a} \Delta^\mu}{\mathcal{Z}(\Lambda_a)} + \mathcal{O}(\Delta^2) \right) \int d\theta p(\theta|D_i,\Lambda_0) \frac{p(\theta|\Lambda_a)^2 + 2 p(\theta|\Lambda_a) \partial_\mu p(\theta|\Lambda)|_{\Lambda_a} \Delta^\mu + \mathcal{O}(\Delta^2)}{p(\theta|\Lambda_0)^2} \nonumber \\
        & - \frac{2m_i^{-1}}{\mathcal{Z}(\Lambda_a)^2} \left( 1 - \frac{\partial_\mu \mathcal{Z}|_{\Lambda_a} \Delta^\mu}{\mathcal{Z}(\Lambda_a)} + \mathcal{O}(\Delta^2) \right) \int d\theta p(\theta|D_i,\Lambda_0) \frac{p(\theta|\Lambda_a)^2 + p(\theta|\Lambda_a)\partial_\mu p(\theta|\Lambda)|_{\Lambda_a} \Delta^\mu + \mathcal{O}(\Delta^2)}{p(\theta|\Lambda_0)^2}
\end{align}
All $\mathcal{O}(1)$ and $\mathcal{O}(\Delta)$ terms cancel.
The leading order term is therefore $\propto \Delta^2 / m_i$, where the proportionality constant is a function of integrals of $p(\theta|\Lambda)$ and its derivatives with respect to $\Lambda$ evaluated at $\Lambda_a$.

Now, the expressions for the selection function are somewhat more complicated algebraically, but they follow the same general form.
Eqs.~\ref{eq:var delta e},~\ref{eq:E expected value}, and~\ref{eq:E cov} produce
\begin{align}
    \mathrm{Var} \left[ \ln \left(\hat{\mathcal{E}}(\Lambda_a) + r^{-1}\right) \right. & \left. - \ln \left(\hat{\mathcal{E}}(\Lambda_b) + r^{-1}\right)\right]_\mathrm{MC} \nonumber \\
        & = \frac{1}{(\mathcal{E}(\Lambda_a) + r^{-1})^2} \left( \frac{P(\mathrm{det}|\mathrm{draw})}{M} \int d\theta p(\theta|\mathrm{det},\mathrm{draw}) \frac{p(\theta|\Lambda_a)^2}{p(\theta|\mathrm{draw})^2} \right) \nonumber \\
        & \quad + \frac{1}{(\mathcal{E}(\Lambda_b) + r^{-1})^2} \left( \frac{P(\mathrm{det}|\mathrm{draw})}{M} \int d\theta p(\theta|\mathrm{det},\mathrm{draw}) \frac{p(\theta|\Lambda_b)^2}{p(\theta|\mathrm{draw})^2} \right) \nonumber \\
        & \quad - \frac{2}{(\mathcal{E}(\Lambda_a) + r^{-1})(\mathcal{E}(\Lambda_b)+r^{-1})} \left( \frac{P(\mathrm{det}|\mathrm{draw})}{M} \int d\theta p(\theta|\mathrm{det},\mathrm{draw}) \frac{p(\theta|\Lambda_a)p(\theta|\Lambda_b)}{p(\theta|\mathrm{draw})^2} \right) \nonumber \\
        & \quad - \frac{1}{M} \left(\frac{\mathcal{E}(\Lambda_a)^2}{(\mathcal{E}(\Lambda_a) + r^{-1})^2}\right) - \frac{1}{M} \left(\frac{\mathcal{E}(\Lambda_b)^2}{(\mathcal{E}(\Lambda_b) + r^{-1})^2}\right) + \frac{2}{M} \left(\frac{\mathcal{E}(\Lambda_a)\mathcal{E}(\Lambda_b)}{(\mathcal{E}(\Lambda_a) + r^{-1})(\mathcal{E}(\Lambda_b)+r^{-1})}\right)
\end{align}
Again, expanding $p(\theta|\Lambda_b) \approx p(\theta|\Lambda_a) + \partial_\mu p(\theta|\Lambda)|_{\Lambda_a} \Delta^\mu$, the first three lines follow exactly the same form as the single-event evidence.
As such, their leading order contribution is $\mathcal{O}(\Delta^2/M)$.
The last three terms simplify to
\begin{align}
    \frac{1}{M} \left( \frac{\mathcal{E}(\Lambda_a)}{\mathcal{E}(\Lambda_a) + r^{-1}} - \frac{\mathcal{E}(\Lambda_b)}{\mathcal{E}(\Lambda_b) + r^{-1}} \right)^2 \approx \frac{1}{M} \left(\frac{r^{-1} \left(\partial_\mu \mathcal{E}|_{\Lambda_a} \Delta^\mu\right)}{(\mathcal{E}(\Lambda_a) + r^{-1})^2} \right)^2
\end{align}
and therefore $\mathrm{Var}[\Delta \ln (\hat{\mathcal{E}}+r^{-1})]_\mathrm{MC} \propto \Delta^2 / M$.
Again, the proportionality constant only depends on integrals of $p(\theta|\Lambda)$ and its derivatives with respect to $\Lambda$ evaluated at $\Lambda_a$.

Combining the fact that each of the variance of each of the $N$ single-event evidences and the selection function all scale as $\mathcal{O}(\Delta^2)$ yields the scaling relation in Eq.~\ref{eq:quadratic scaling}.


\section{Marginalization over Monte Carlo Uncertainty in $\mathcal{Z}_i$}
\label{sec:characteristic function}

We first consider the distribution of Monte Carlo sums analytically, beginning with the basic Monte Carlo estimator
\begin{equation}
    \hat{f} \equiv \frac{1}{m}\sum\limits_k^m f(\theta_k)
\end{equation}
where we have $m$ samples drawn from
\begin{equation}
    \theta_k \sim p(\theta)
\end{equation}
and we approximate the integral
\begin{equation}
    F \equiv \int d\theta\, p(\theta) f(\theta)
\end{equation}
Now, we can write the measure for $\hat{f}|p(\theta),m$ as
\begin{equation}
    p(\hat{f}|p(\theta), m) = \int \left[ \left(\prod\limits_j^m d\theta_j\, p(\theta_j) \right) \delta\left(\hat{f} - \frac{1}{m}\sum\limits_k^m f(\theta_k) \right) \right]
\end{equation}
However, we find it easier to deal with the corresponding characteristic function
\begin{align}
    \phi_{\hat{f}|p(\theta),m}(s) \equiv \mathrm{E}[e^{i s \hat{f}}]_{\hat{f}|F,m}
        & = \int d\hat{f} e^{i s \hat{f}} \int \left[ \left(\prod\limits_j^m d\theta_j\, p(\theta_j) \right)  \delta\left(\hat{f} - \frac{1}{m}\sum\limits_k^m f(\theta_k) \right) \right] \nonumber \\
        & = \int \left[ \left(\prod\limits_j^m d\theta_j\, p(\theta_j) \right) e^{(i s/m) \sum\limits_k^m f(\theta_k)} \right]\nonumber \\
        & = \left[ \int d\theta\, p(\theta) e^{i s f(\theta) / m} \right]^m \nonumber \\
        & = e^{i s F} \left[ \int d\theta\, p(\theta) e^{i s (f(\theta) - F)/ m} \right]^m
\end{align}
and then express
\begin{equation}
    p(\hat{f}|F, \mathrm{HM},m) = \int ds\, e^{-i s \hat{f}} \phi_{\hat{f}|p(\theta),m}(s)
\end{equation}
where we have split the dependence on $p(\theta)$ into the mean ($F$) and higher (central) moments ($\mathrm{HM}$).\footnote{
Even though $F$ is completely determined by $p(\theta)$, we divide up the dependence of $\hat{f}$ into the first moment $F$ and the higher moments (also determined by $p(\theta)$) in order to compute a distribution for $F$.
}

Furthermore, we are interested in constructing a distribution for $F$ given $\hat{f}$, $\mathrm{HM}$, and $m$.
We can do this via Bayes theorem as follows
\begin{equation}
    p(F|\hat{f},\mathrm{HM},m) \propto p(\hat{f}|F,\mathrm{HM},m) p(F|\mathrm{HM},m)
\end{equation}
with the corresponding characteristic function
\begin{align}
    \phi_{F|\hat{f},\mathrm{HM},m}(r) \equiv \mathrm{E}[e^{i r F}]_{F|\hat{f},\mathrm{HM},m}
        & \propto \int dF\, e^{i r F} p(F|\mathrm{HM}) p(\hat{f}|F,\mathrm{HM},m) \nonumber \\
        & \propto \int dF\, e^{i r F} p(F|\mathrm{HM}) \int ds\, e^{-i s \hat{f}} \left( e^{isF} \left[ \int d\theta\, p(\theta) e^{is(f(\theta)-F)/m} \right]^m \right) \nonumber \\
        & \propto \int ds\, e^{-i s \hat{f}} \left[ \int d\theta\, p(\theta) e^{is(f(\theta)-F)/m} \right]^m \int dF\, e^{i (r+s) F} p(F|\mathrm{HM})
\end{align}
where, in the last line we can switch the order of integration because
\begin{equation}
    \int d\theta\, p(\theta) e^{is(f(\theta)-F)/m} \nonumber
\end{equation}
only depends on higher (central) moments of $f(\theta)$, which are independent of $F$.
This allows us to factor it outside the integral over $F$.

To that end, because the higher moments of $f(\theta)$ are independent of $F$, we approximate $p(F|\mathrm{HM}) = p(F)$ as a uniform distribution support over the whole real line.
We can then further simplify this expression to
\begin{align}
    \phi_{F|\hat{f},\mathrm{HM},m}(r)
        & \propto \int ds\, e^{-is\hat{f}} \left[ \int d\theta\, p(\theta) e^{is(f(\theta)-F)/m} \right]^m \delta(r+s) \nonumber \\
        & = e^{ir\hat{f}} \left[ \int d\theta\, p(\theta) e^{ir(F-f(\theta))/m} \right]^m
\end{align}
with the proportionality constant determined by the condition that $\phi(r=0) = 1$.
With this in hand, we can evaluate moments of $F|\hat{f},\mathrm{HM},m$ as needed in what follows via
\begin{equation}
    \mathrm{E}[F^n]_{F|\hat{f},\mathrm{HM},m} = \left. \frac{\partial^n}{(i\partial r)^n} \phi_{F|\hat{f},\mathrm{HM},m} \right|_{r=0}
\end{equation}

Now consider the marginalization over $\mathcal{Z}_i$.
In this case, we obtain the desired expectation value with the association $F, \hat{f} \rightarrow \mathcal{Z}, \hat{\mathcal{Z}}$ and
\begin{align}
    \left. \frac{\partial}{i\partial r} \phi_{F|\hat{f},\mathrm{HM},m}(r) \right|_{r=0}
        & = \left. \left( \hat{f} + m \frac{\int d\theta\, p(\theta) \left(\frac{F-f(\theta)}{m}\right) e^{ir(F-f(\theta))/m}}{\int d\theta\, p(\theta) e^{ir(F-f(\theta))/m}} \right) \phi_{F|\hat{f},\mathrm{HM},m}(r) \right|_{r=0} \nonumber \\
    \mathrm{E}[F]_{F|\hat{f},\mathrm{HM},m}
        & = \hat{f}
\end{align}
implying $\mathrm{E}[\mathcal{Z}_i]_{\mathcal{Z}_i|\hat{\mathcal{Z}}_i,m_i} = \hat{\mathcal{Z}}_i$.

\end{document}